\documentclass[lettersize,journal]{IEEEtran}
\usepackage{amsmath,amsfonts}
\usepackage{algorithmic}
\usepackage{algorithm}
\usepackage{array}
\usepackage[caption=false,font=normalsize,labelfont=sf,textfont=sf]{subfig}
\usepackage{textcomp}
\usepackage{stfloats}
\usepackage{url}
\usepackage{verbatim}
\usepackage{graphicx}
\usepackage{cite}
\usepackage{xcolor,hyperref}
\usepackage{makecell}
\hypersetup{colorlinks = true,linkcolor= blue,citecolor=red}
\usepackage{pifont}
\newcommand{\cmark}{\ding{51}}  
\newcommand{\xmark}{\ding{55}}  
\usepackage{rotating}
\usepackage{booktabs}
\usepackage{amssymb}
\usepackage{pifont}

\begin{document}

\makeatletter
\def\ps@IEEEtitlepagestyle{%
  \def\@oddfoot{\mycopyrightnotice}%
  \def\@evenfoot{}}
\def\mycopyrightnotice{%
  \hfill \footnotesize
  Accepted in: \textit{IEEE Open Journal of the Communications Society (OJ-COMS)}, Oct. 27, 2025. \hfill
}
\makeatother

\title{Cooperative NOMA Meets Emerging Technologies: A Survey for Next-Generation Wireless Networks}

\author{Mahmoud M. Salim, Suhail I. Al-Dharrab, \IEEEmembership{Senior Member, IEEE}, Daniel Benevides Da Costa, \IEEEmembership{Senior Member, IEEE}, and Ali H. Muqaibel,
\IEEEmembership{Senior Member, IEEE}
\thanks{All authors are with the Center for Communications Systems and Sensing, King Fahd University of Petroleum $\&$ Minerals, Dhahran 31261, Saudi Arabia (E-mails: $\{$mahmoud.elemam, suhaild$\}$@kfupm.edu.sa; danielbcosta@ieee.org; muqaibel@kfupm.edu.sa). Also, Suhail I. Al-Dharrab, Daniel Benevides Da Costa, and Ali H. Muqaibel are with the Electrical Engineering Department, King Fahd University of Petroleum  $\&$ Minerals, Dhahran 31261, Saudi Arabia. Mahmoud M. Salim is the corresponding author.

}}

\markboth{Journal of \LaTeX\ Class Files}%
{Shell \MakeLowercase{\textit{et al.}}: A Sample Article Using IEEEtran.cls for IEEE Journals}

\maketitle

\begin{abstract}
The emerging demands of sixth-generation wireless networks, such as ultra-connectivity, native intelligence, and cross-domain convergence, are bringing renewed focus to cooperative non-orthogonal multiple access (C-NOMA) as a fundamental enabler of scalable, efficient, and intelligent communication systems. C-NOMA builds on the core benefits of NOMA by leveraging user cooperation and relay strategies to enhance spectral efficiency, coverage, and energy performance. This article presents a unified and forward-looking survey on the integration of C-NOMA with key enabling technologies, including radio frequency energy harvesting, cognitive radio networks, reconfigurable intelligent surfaces, space-air-ground integrated networks, and integrated sensing and communication-assisted semantic communication. Foundational principles and relaying protocols are first introduced to establish the technical relevance of C-NOMA. Then, a focused investigation is conducted into protocol-level synergies, architectural models, and deployment strategies across these technologies. Beyond integration, this article emphasizes the orchestration of C-NOMA across future application domains such as digital twins, extended reality, and e-health. In addition, it provides an extensive and in-depth review of recent literature, categorized by relaying schemes, system models, performance metrics, and optimization paradigms, including model-based, heuristic, and AI-driven approaches. Finally, open challenges and future research directions are outlined, spanning standardization, security, and cross-layer design, positioning C-NOMA as a key pillar of intelligent next-generation network architectures.

\end{abstract}

\IEEEpeerreviewmaketitle

\section{Introduction}\label{Sec_Intro}
\IEEEPARstart{N}{ext}-generation networks are set to revolutionize various sectors through innovative applications such as cognitive cities, immersive extended reality (XR), and vehicle-to-everything (V2X) communications \cite{nguyen-2022,road-wang-2023, survey-de-alwis-2021, comprehensive-jiang-2021}. Powered by cutting-edge next-generation technologies, these applications aim to deliver an exceptional quality of experience (QoE) through ultra-high data rates, ultra-low latency, and massive connectivity, addressing the explosive growth of mobile devices \cite{barakabitze2019qoe,zhao2023qoe,chen2023joint,pimpalkar2024novel,henrique20236g}. An expected crucial enabler is cooperative non-orthogonal multiple access (C-NOMA), which uses relays to enhance signal strength and coverage \cite{interplay-vaezi-2019,rateoptimal-maraqa-2020}. Those relays are expected to be full-duplex (FD) to align with the next-generation requirements \cite{fullduplex-mohammadi-2023}. Besides, it allows multiple users to share the same frequency band through power-level signal superposition, therefore improving spectrum efficiency (SE) and communication reliability. C-NOMA can be integrated with energy harvesting (EH) technologies, space-air-ground integrated networks (SAGINs), cognitive radio (CR) networks, semantic communication, integrated sensing and communication (ISAC), haptic communications and digital twins, and reconfigurable intelligent surfaces (RISs) \cite{salim2025snr,10038657, 8368236,8972429,smart-gong-2020}. This integration is envisioned to greatly enhance energy efficiency (EE), scalability, spectrum utilization, and overall network performance. Nonetheless, significant challenges remain in the standardization and practical modeling of emerging technologies, requiring continued research efforts to bridge theoretical advances with real-world deployments \cite{vardhan2025aber,lin2025bridge}. 

EH enhances the efficiency and sustainability of C-NOMA networks by enabling nodes to collect energy from radio frequency (RF) signals and renewable energy sources such as solar, or wind \cite{salim-2020}. However, the intermittent power supply from renewable sources and the modest energy captured from RF signals \cite{salim2022rf,salim2023rf} is challenging. It is expected that RF EH is more feasible when combined with beamsteering and beamforming technologies in multiple-input multiple-output (MIMO) systems, boosting the RF amount harvested at the receiver. On the other hand, developing sophisticated energy management techniques and power allocation strategies is crucial to dealing with renewable energy's unpredictable and intermittent nature. SAGINs offer adaptability and scalability for C-NOMA networks, especially in rapid deployment scenarios, but face challenges in maintaining stable communication and managing diverse components. Integrating C-NOMA with CR networks improves spectrum utilization by exploiting underused frequency bands through dynamic access approaches. This integration enhances network performance but introduces complexity in spectrum management and potential security issues. On the other hand, RISs can transform signal propagation and coverage in C-NOMA systems by controlling radio wave reflection, leading to better efficiency and user experience. Challenges include the complexity of RIS deployment, real-time optimization, and cost concerns.

Combining C-NOMA with semantic communication and ISAC advances next-generation networks by enhancing data processing and environmental awareness \cite{10437438}. Semantic communication uses deep neural networks (DNNs) to improve transmission efficiency by focusing on essential data features, while ISAC integrates sensing and communication for real-time awareness. Despite the benefits, challenges include the developing stage of semantic communication, the complexity of integrating C-NOMA with ISAC for simultaneous tasks, and issues with interoperability and standardization. Advanced artificial intelligence (AI) techniques, such as generative AI, native AI, and deep reinforcement learning (DRL), can play a pivotal role in addressing these challenges by creating synthetic data to enhance the training of semantic communication models and improving feature extraction and modulation techniques \cite{10398474}. It can also generate adaptive models for simultaneous communication and sensing tasks, helping to streamline the integration of C-NOMA with ISAC. 

\subsection{Motivation and Contributions}

Despite its demonstrated SE and support for massive connectivity, NOMA was never formally adopted into 3GPP’s fifth-generation (5G) new radio (NR) standards, remaining instead at the level of a technical report in Release 15 \cite{3gpp-tr-38.812}. As research agendas have shifted toward emerging paradigms such as ISAC, RIS, and satellite-terrestrial convergence, NOMA appears to have lost traction in next-generation standardization discussions. However, the evolving demands of 6G, namely ultra-connectivity, intelligence, and convergence, align well with NOMA’s inherent strengths. This opens an opportunity for a “second chance” for NOMA, particularly in cooperative forms, i.e., C-NOMA, when tightly integrated with these emerging technologies. The synergies revealed in recent studies suggest that C-NOMA could play a pivotal role in shaping intelligent and spectrum-efficient communication architectures in 6G and beyond.

\begin{figure*}[th]
\centering
\centerline{\includegraphics[width=2\columnwidth]{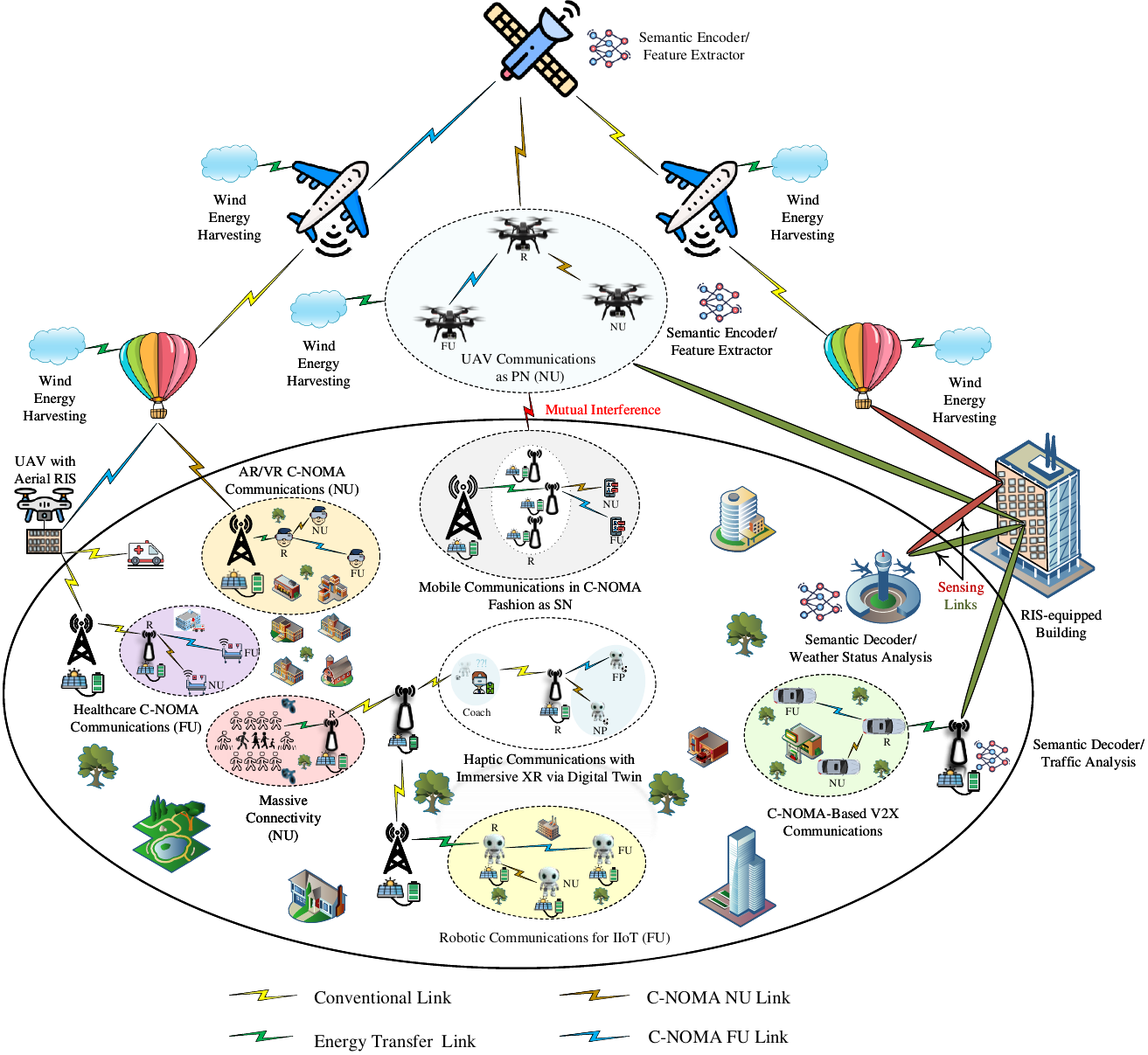}}
\caption{Envisioned future next-generation networks vision with C-NOMA.}
\label{fig_Big_Pict}
\end{figure*}

Building on this renewed relevance, the fusion of immersive XR, digital twins, haptic communication, and C-NOMA defines a novel paradigm in communication systems. This integration enables real-time, immersive experiences with enhanced SE and ultra-low latency, supporting precise remote control in diverse applications such as healthcare, industrial automation, and sports. To meet the stringent demands of these applications, this article explores how a suite of enabling technologies, including EH, RISs, semantic communication, and SAGINs, can be synergistically integrated with C-NOMA. AI plays a central role in orchestrating this ecosystem, optimizing performance, managing complexity, and enabling intelligent adaptation across dynamic environments \cite{he2024nuclear}. Figure~\ref{fig_Big_Pict} illustrates the envisioned convergence of C-NOMA with these cutting-edge technologies, emphasizing its transformative potential for next-generation networks.

Table~\ref{tab:compact_comparison} presents a comparative analysis of our survey against recent and representative works in the literature. As illustrated, while prior surveys have addressed specific aspects of NOMA, such as RIS-assisted designs, ISAC integration, or UAV-based applications, none of them has comprehensively examined the role of C-NOMA across the spectrum of key enabling technologies envisioned for 6G networks. Our work distinctly contributes by unifying the treatment of C-NOMA with RIS, ISAC, EH, CR, SAGINs, and semantic communication under a single framework. Moreover, we introduce a dedicated taxonomy and evaluation of optimization strategies (including model-based, heuristic, and AI-driven approaches) specific to C-NOMA deployments. In contrast to previous works that primarily focus on physical layer configurations, we emphasize the orchestration of C-NOMA across intelligent applications, such as digital twins, XR, and e-health, highlighting its potential for multi-dimensional service provisioning. As the table reveals, this work is the first to offer a holistic, application-aware, and technology-integrated survey on C-NOMA for future intelligent and sustainable wireless networks.

\begin{table*}[th]
\centering
\caption{Comparison of our survey with existing survey papers and magazine articles.}
\begin{tabular}{|p{1.5cm}|c|c|c|c|c|c|c|c|c|c|}
 \hline
\textbf{Ref.} & \textbf{Year} & \textbf{Type} & \textbf{C-NOMA} & \textbf{RIS} & \textbf{ISAC} & \textbf{EH} & \textbf{SAGIN} & \textbf{CR} & \textbf{Semantic Communications} & \textbf{AI/Optimization} \\ \hline
\cite{islam-2016-noma} & 2016 & Survey & \cmark & \xmark & \xmark & \cmark & \xmark & \xmark & \xmark & \xmark \\ \hline
\cite{ding-2017-survey} & 2017 & Survey & \cmark & * & * & * & * & \xmark & \xmark & * \\ \hline
\cite{dai-2018-survey} & 2018 & Survey & \cmark & * & \xmark & \xmark & \xmark & \xmark & \xmark & * \\ \hline
\cite{shahab-2020-grantfree} & 2020 & Survey & \cmark & \xmark & \xmark & \xmark & \xmark & \xmark & \xmark & \cmark \\ \hline
\cite{maraqa-2020-rateoptimal} & 2020 & Survey & \cmark & \cmark & * & \cmark & * & * & \xmark & \cmark \\ \hline
\cite{liu-2024-ris} & 2024 & Survey & * & \cmark & * & * & \cmark & * & \xmark & \cmark \\ \hline
\cite{ahmed-2024-unveiling} & 2024 & Survey & * & \cmark & \cmark & \cmark & \cmark & \cmark & * & \cmark \\\hline
\cite{sharma-2024-irs} & 2024 & Survey & \cmark & \cmark & * & \xmark & * & \xmark & \xmark & * \\\hline
\cite{sur-2024-survey} & 2024 & Survey & * & \cmark & \cmark & \xmark & \xmark & \xmark & \xmark & * \\\hline
\cite{budhiraja-2021-systematic} & 2021 & Survey & * & * & \xmark & * & \xmark & \xmark & \xmark & * \\\hline
\cite{vaezi-2019-interplay} & 2019 & Survey & * & \cmark & * & \cmark & * & \cmark & \xmark & \xmark \\\hline
\cite{dai-2018-survey} & 2018 & Survey & \xmark & \xmark & \xmark & \xmark & \xmark & \xmark & \xmark & \xmark \\\hline
\cite{mayarakaca-2024-survey} & 2024 & Survey & * & \xmark & \xmark & * & \cmark & \xmark & \xmark & \cmark \\\hline
\cite{sarkar-2024-comprehensive} & 2024 & Survey & \xmark & \cmark & \xmark & * & * & \xmark & \xmark & \cmark \\\hline
\cite{pakravan-2023-security} & 2023 & Survey & \cmark & * & \xmark & \xmark & \xmark & \xmark & \xmark & \xmark \\\hline
\cite{ahmed-2024-backscatter} & 2024 & Survey & \xmark & * & \xmark & * & * & \xmark & \xmark & \xmark \\ \hline
\cite{raza-2023-ai} & 2023 & Survey & \xmark & \xmark & \xmark & \xmark & \xmark & \xmark & \xmark & \cmark \\ \hline
\textbf{This work} & 2024 & Survey & \cmark & \cmark & \cmark & \cmark & \cmark & \cmark & \cmark & \cmark \\ 
\hline

\multicolumn{11}{l}{\textit{[\cmark: Thoroughly discussed; *: Partially; \xmark: Not addressed]}
} \\ 

\bottomrule
\end{tabular}
\label{tab:compact_comparison}
\end{table*}

In summary, the key contributions of this article, which presents a unified and forward-looking perspective on the integration of C-NOMA with emerging technologies for future wireless networks, are as follows:
\begin{itemize}
\item The article presents essential foundations, including the basic principles of C-NOMA and its relaying protocols, which serve as the basis for understanding its performance advantages and practical relevance in next-generation wireless systems.

\item A focused investigation is conducted on the architectural and functional integration of C-NOMA with key enabling technologies, namely RF EH, CR networks, RIS, SAGINs, and ISAC-assisted semantic communication. This includes protocol-level interactions, deployment strategies, and enabling mechanisms. Furthermore, the orchestration across associated application domains is examined to illustrate how C-NOMA can adapt to diverse and dynamic real-world scenarios.

\item An extensive survey of recent literature is performed on the integration of C-NOMA with the aforementioned key enabling technologies. Each domain is examined in terms of system models, relaying strategies, performance metrics, and optimization techniques. A dedicated subsection is also provided to categorize optimization approaches in C-NOMA networks into model-based, heuristic, and AI-driven methods. Besides, summary tables are outlined to capture representative contributions, support comparative evaluation, and guide future research.

\item A detailed analysis of technical challenges and open research problems is presented, organized by technology dimension: C-NOMA vs EH, RIS, ISAC-assisted semantic communication, SAGINs, digital twins, quantum communication, and security and privacy. In addition, the article outlines standardization challenges and identifies critical research opportunities for the integration of C-NOMA in 6G systems.
\end{itemize}

\subsection{Paper Organization}
The remainder of this article is organized as follows. Section~\ref{Sec_Foundation} provides the essential foundations, including the basic principles of C-NOMA and its relaying protocols. In Section~\ref{Sec_Integration}, a focused investigation is carried out on the architectural and functional integration of C-NOMA with key enabling technologies such as RF EH, CR networks, RIS, SAGINs, and ISAC-assisted semantic communication, along with their orchestration across envisioned application domains. Section~\ref{Sec_RW} presents an extensive review of recent literature on the integration of C-NOMA with these technologies, including a dedicated subsection on optimization approaches categorized as model-based, heuristic, and AI-driven. Section~\ref{Sec_chall} outlines the critical challenges and open research problems for C-NOMA in the context of future wireless networks. Finally, the paper is concluded in Section~\ref{Sec_Conc}. For reference, the most common abbreviations used throughout the article are listed in Table~\ref{tab:Acronyms}.

\begin{figure}[h!]
\centering
\centerline{\includegraphics[width=\columnwidth]{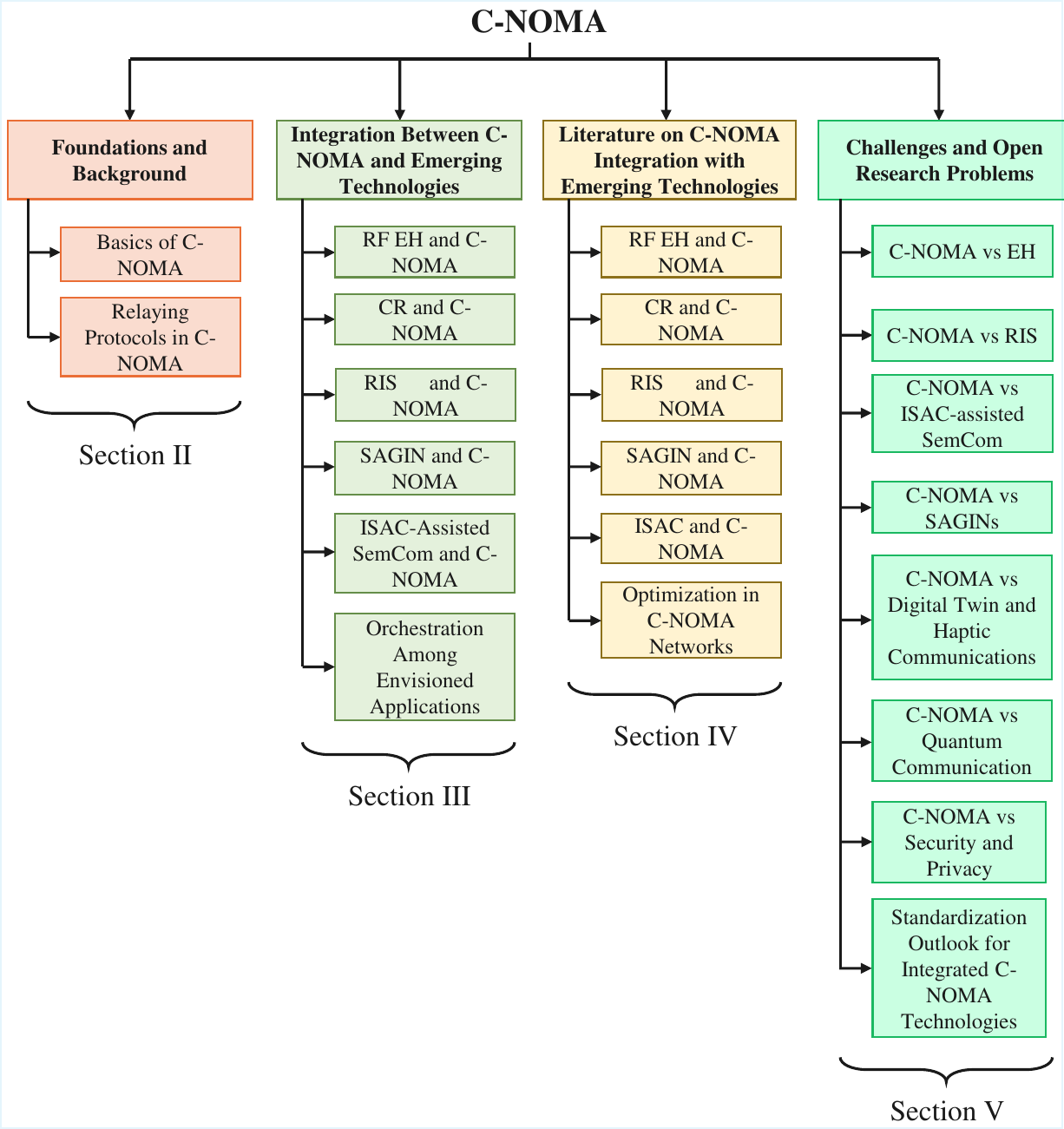}}
\caption{Paper Organization.}
\label{fig_Relaying_FD}
\end{figure}

\begin{table*}[!t]
\caption{List of Acronyms}
\label{Tab:Acronyms}
\centering
\renewcommand{\arraystretch}{1.0}
\setlength{\tabcolsep}{6pt}
\begin{tabular}{|l|l|l|l|}
\hline
\textbf{Acronym} & \textbf{Definition} & \textbf{Acronym} & \textbf{Definition} \\
\hline
6G & Sixth generation & AF & Amplify-and-forward \\ \hline
ADC & Analog-to-digital converters & AI & Artificial intelligence \\ \hline
AoI & Age of information & AO & Alternating optimization \\ \hline
AR & Augmented reality & BER & Bit error rate \\ \hline
BS & Base station & C-NOMA & Cooperative NOMA \\ \hline
CPU & Central processing unit & CR & Cognitive radio \\ \hline
DF & Decode-and-forward & DDPG & Deep deterministic policy gradient \\ \hline
DNNs & Deep neural networks & DQN & Deep Q-networks \\ \hline
DRL & Deep reinforcement learning & EE & Energy efficiency \\ \hline
EH & Energy harvesting & EHR & Energy harvesting relay \\ \hline
FD & Full-duplex & FP & Fractional programming \\ \hline
FU & Far user & GA & Genetic algorithm \\ \hline
HDAF & Hybrid decode- and amplify-and-forward & HI & Hardware impairment \\ \hline
HTSPS & Hybrid TS and PS & IIoT & Industrial IoT \\ \hline
IoT & Internet-of-Things & ISAC & Integrated sensing and communication \\ \hline
LEO & Low earth orbit & MAC & Medium access control \\ \hline
MEC & Mobile edge computing & MIMO & Multiple-input multiple-output \\ \hline
MINLP & Mixed-integer nonlinear programming & MISO & Multiple-input single-output \\ \hline
MM & Majorization-minimization & MRC & Maximum ratio combining \\ \hline
NOMA & Non-orthogonal multiple access & NU & Near user \\ \hline
NV & Near vehicles & OMA & Orthogonal multiple access \\ \hline
OP & OP & PR & Primary receiver \\ \hline
PS & Power splitting & PSO & Particle swarm algorithm \\ \hline
PT & Primary transmitter & PU & Primary user \\ \hline
QPSO & Quantum-behaved particle swarm optimization & QoE & Quality of experience \\ \hline
QoS & Quality-of-service & RE & Renewable energy \\ \hline
RF & Radio frequency & RISs & Reconfigurable intelligent surfaces \\ \hline
RSMA & Rate-splitting multiple access & RSU & Roadside unit \\ \hline
RV & Relay vehicle & SAGINs & Space-air-ground integrated networks \\ \hline
SCA & Successive convex approximation & SDR & Semi-definite relaxation \\ \hline
SE & Spectral efficiency & SIC & Successive interference cancellation \\ \hline
SN & Secondary network & SINR & Signal-to-interference-plus-noise ratio \\ \hline
STAR-RISs & Simultaneously Transmitting and Reflecting RISs & SNR & Signal-to-noise ratio \\ \hline
SU & Secondary users & STINs & Satellite-terrestrial integrated networks \\ \hline
TS & Time-switching & SWIPT & Simultaneous wireless information and power transfer \\ \hline
UAVs & Unmanned aerial vehicles & U2U & User-to-user \\ \hline
VR & Virtual reality & V2X & Vehicle-to-everything \\ \hline
XR & Extended reality & WPT & Wireless power transfer \\ \hline
\end{tabular}
\label{tab:Acronyms}
\end{table*}

\section{Foundations and Background}
\label{Sec_Foundation}
\subsection{Basics of C-NOMA
}\label{SubSec_CNOMA}
 Power-domain NOMA\footnote{Hereafter, NOMA means power-domain NOMA} enhances SE by allowing multiple users, typically a near user (NU) and a far user (FU), to share the same frequency band by superimposing their signals at different power levels \cite{islam-2016,ding-2017,dai-2015}. This distinguishes NOMA from traditional orthogonal multiple access (OMA) methods. C-NOMA is an advanced version of NOMA, which introduces relaying to improve performance further. Two common configurations are used in C-NOMA. The first involves the NU relaying the FU data \cite{liu-2022-NOMA,chaki-2023}, while the second places a dedicated relay between the transmitter and the NU and the FU \cite{liu-2021-NOMA, yeom-2022,arzykulov-2021}. This second setup enhances reliability and coverage, making it suitable for ultra-dense networks. While it improves power management, potential gains in spectral or EE depend on specific configurations and trade-offs. As such, it forms the main focus of our work.  
 
\subsection{Relaying Protocols in C-NOMA}
\begin{figure}[h!]
\centering
\centerline{\includegraphics[width=0.9\columnwidth]{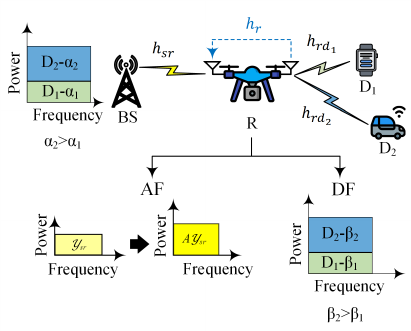}}
\caption{A C-NOMA system model to explain the different relaying protocols.}
\label{fig_Relaying_FD}
\end{figure}
Different relaying protocols are adopted in C-NOMA communications to boost the system performance. According to the literature, these protocols can be divided into three types i.e., amplify-and-forward (AF), decode-and-forward (DF), and hybrid decode- and amplify-and-forward (HDAF)\cite{liu-2016,li-2021,bai-2015,salim-2020}. We may understand the main idea of each protocol according to the terrestrial-UAV communication model in Fig. \ref{fig_Relaying_FD}. The model depicts a base station (BS) $S$ that aims to transmit the superimposed message $x_s= \sum_{i=1}^{2} \sqrt{\alpha_i} x_{i}$ to two remote IoT devices namely $D_1$ and $D_2$ to decode their messages $x_1$ and $x_2$, respectively. Notably, at the BS, $\alpha_1$ and $\alpha_2$ are the power coefficients for $D_1$ and $D_2$, respectively; $\alpha_i \in [0,1]$ and $\sum_{i=1}^{2} \alpha_i=1$, $\forall{i=\{1,2\}}$. In practical scenarios, there could be high attenuation or blockage between $S$ on one side and $D_1$ and $D_2$ on the other side. Therefore, an intermediate relay $R$ is utilized.
\subsubsection{Amplify-and-Forward}
In the light of the C-NOMA scenario in Fig. \ref{fig_Relaying_FD}, the AF protocol offers a simple method where $R$ amplifies the superimposed signal received from $S$ and transmits it to $D_1$ and $D_2$ without decoding. The received message at $D_i, \forall{i\in\{1,2\}}$, is $y^{AF}_{d_i}=A (h_{rd_i}\sqrt{P_r}y_{sr}+h_r\sqrt{P_r}s_r)+n_{d_i}$, where $y_{sr}$ is the received signal at $R$ and $A$ is the amplification factor. $h_{sr}$ is the channel gain between $S$ and $R$, while $h_{rd_i}$ is that between $R$ and $D_i$. Also, $P_r$ is the relay transmit power, and $h_r\sqrt{P_r}x_r$ represents the relay's self-interference component. Besides, $n_{d_i}\sim \mathcal{CN}(0,\sigma^2_{d_i})$ denoting the additive white Gaussian noise at the receiver $D_i$. Since no successive interference cancellation (SIC) is needed for decoding, AF is considered a simple scheme that requires less processing power and is appropriate for situations where the signal-to-noise ratio (SNR) is low to moderate \cite{singh-2021}. However, AF suffers from noise amplification and inefficient resource utilization. Moreover, it could be vulnerable to channel impairments and interference, which could lower the overall performance and signal quality \cite{do-2021}.

Despite the absence of decoding at the relay, SIC is still employed at the receivers to extract the intended signals. Specifically, the NU $D_1$ first decodes the stronger signal intended for the FU $D_2$, cancels its effect, and then decodes its own signal. Conversely, $D_2$, benefiting from a higher power allocation, directly decodes its message. This division of decoding tasks maintains the core principle of C-NOMA. However, this arrangement brings additional challenges. The amplified noise introduced by the relay can degrade signal quality, complicating the SIC process, especially under low SNR or highly dynamic channel conditions. Moreover, the entire SIC computational burden falls on user devices, which may be a limiting factor for low-power IoT devices. These limitations underscore the trade-off between the simplicity of AF relaying and the practical complexities of SIC at the receiver end \cite{ezhilazhagan-2022}.

\subsubsection{Decode-and-Forward}
The DF protocol can be explained for C-NOMA scenarios through Fig. \ref{fig_Relaying_FD}. It involves that the relay $R$ decodes the received composite signal from $S$ to extract messages $x_1$ and $x_2$. Before sending the messages to the $D_1$ and $D_2$, the relay first re-encodes them using the power factors $\beta_1$ and $\beta_2$ following the NOMA principle \cite{cooperative-ding-2015,
survey-ding-2017}; $\beta_i \in [0,1]$ and $\sum_{i=1}^{2} \beta_i=1$, $\forall{i=\{1,2\}}$. The received signal at $D_i$ is represented as $y^{DF}_{d_i}=h_{rd_i}\sqrt{P_r}x_{rd_i}+n_{d_i}$, in which $x_{rd_i}=\sum_{i=1}^{2} \sqrt{\beta_i} x_{i}$ denotes the transmitted signal of the relay, and $n_{d_i}\sim \mathcal{CN}(0,\sigma^2_{d_i})$ signifies the additive white Gaussian noise at receiver $D_i$. Compared to AF, DF provides better error performance and dependability, but it has more complexity and processing latency, which can be problematic, especially for real-time applications. Furthermore, in situations when there is a lot of interference or significant channel impairments, the advantages of DF might be limited. Last but not least, DF has difficulties with SIC and self-interference impacting the decoding process of the received messages \cite{li-2020}.

To recover the individual messages, the relay performs SIC by first decoding the stronger signal intended for the NU $D_1$, removing it from the superimposed transmission, and then decoding the weaker signal meant for the FU $D_2$. After decoding, the relay re-encodes the messages using NOMA power coefficients and forwards them to the users. This enables better system error performance and robustness in the presence of interference and challenging channel conditions \cite{wu-2021}. Nonetheless, this gain in reliability comes at the cost of increased processing complexity and latency, as the relay bears the computational load of SIC. These drawbacks can be critical in real-time or delay-sensitive scenarios. Moreover, under full-duplex operation, the presence of self-interference at the relay complicates the decoding process. Effective SIC using advanced signal processing becomes essential to maintain SIC accuracy and prevent error propagation \cite{fidan-2021}.
\subsubsection{Hybrid Decode- and Amplify-and-Forward}
The HDAF protocol functions as a flexible relaying method in C-NOMA scenarios. It modifies its operation dynamically according to the relay's decoding capability and the state of the channel \cite{ramesh-2023}. HDAF switches between the AF and DF modes based on the signal-to-interference-plus-noise ratio (SINR) threshold. In particular, HDAF smoothly switches to the AF mode when the SINR at the relay drops below a predetermined threshold i.e., {\footnotesize$\gamma_r=\frac{|h_{sr}|^2P_s}{|h_{r}|^2P_r+\sigma_{r}^2}\leq \gamma_{min}$}, indicating unreliable decoding of the forwarded signal. On the other hand, HDAF uses the DF protocol if $\gamma_r>\gamma_{min}$, indicating successful decoding at the relay. With this flexible strategy, resource usage is maximized, and reliable performance is guaranteed in a variety of channel situations \cite{salem-2021}.
 
Since HDAF alternates between DF and AF based on the relay’s SINR, the role and placement of SIC depend on the operating mode. When DF is selected, the relay decodes the composite signal and performs SIC, while in AF mode, SIC is pushed to the receivers. This adaptive approach enhances flexibility and robustness across diverse channel conditions, making HDAF especially well-suited for sixth-generation (6G) scenarios with dynamic and heterogeneous links \cite{salem-2021}. However, the joint optimization of power allocation and switching thresholds introduces notable complexity. Smooth mode transitioning is crucial to avoid performance drops, especially under rapidly varying channels. Additionally, implementing real-time switching and managing SIC location shifts adds to the computational burden and implementation cost \cite{singh-2022-Vibhum}.

Beyond relaying mode selection, UAV-mounted RIS trajectory significantly impacts HDAF-enabled C-NOMA performance. Optimal placement improves SINR and influences AF/DF switching. Density-aware and Fermat point-based methods \cite{lyu2021fast,dudik-2015} can guide UAV positioning, either minimizing squared or total Euclidean distances to users, often prioritizing FUs to enhance SIC reliability. However, UAV mobility is constrained by factors such as velocity, endurance, and stability, requiring trajectory planning that balances communication gains with flight limitations.

 Deep learning, especially DRL and generative AI solutions \cite{drl-ou-2023, noma-li-2020, drl-shahjalal-2024, drl-hu-2024, offloading-zhu-2022}, can play a significant role in optimizing power allocation and user pairing, further enhancing spectrum management and maximizing the performance of these relaying configurations. This configuration offers a robust solution for next-generation applications like immersive XR and semantic communication, delivering high-performance communication \cite{diallo-2024,wang-2025, wang-2025-2}.

\section{Integration Between C-NOMA and Emerging Technologies} \label{Sec_Integration}
\subsection{RF Energy Harvesting and C-NOMA}
\begin{figure*}[!t]
\centering
\centerline{\includegraphics[width=2\columnwidth]{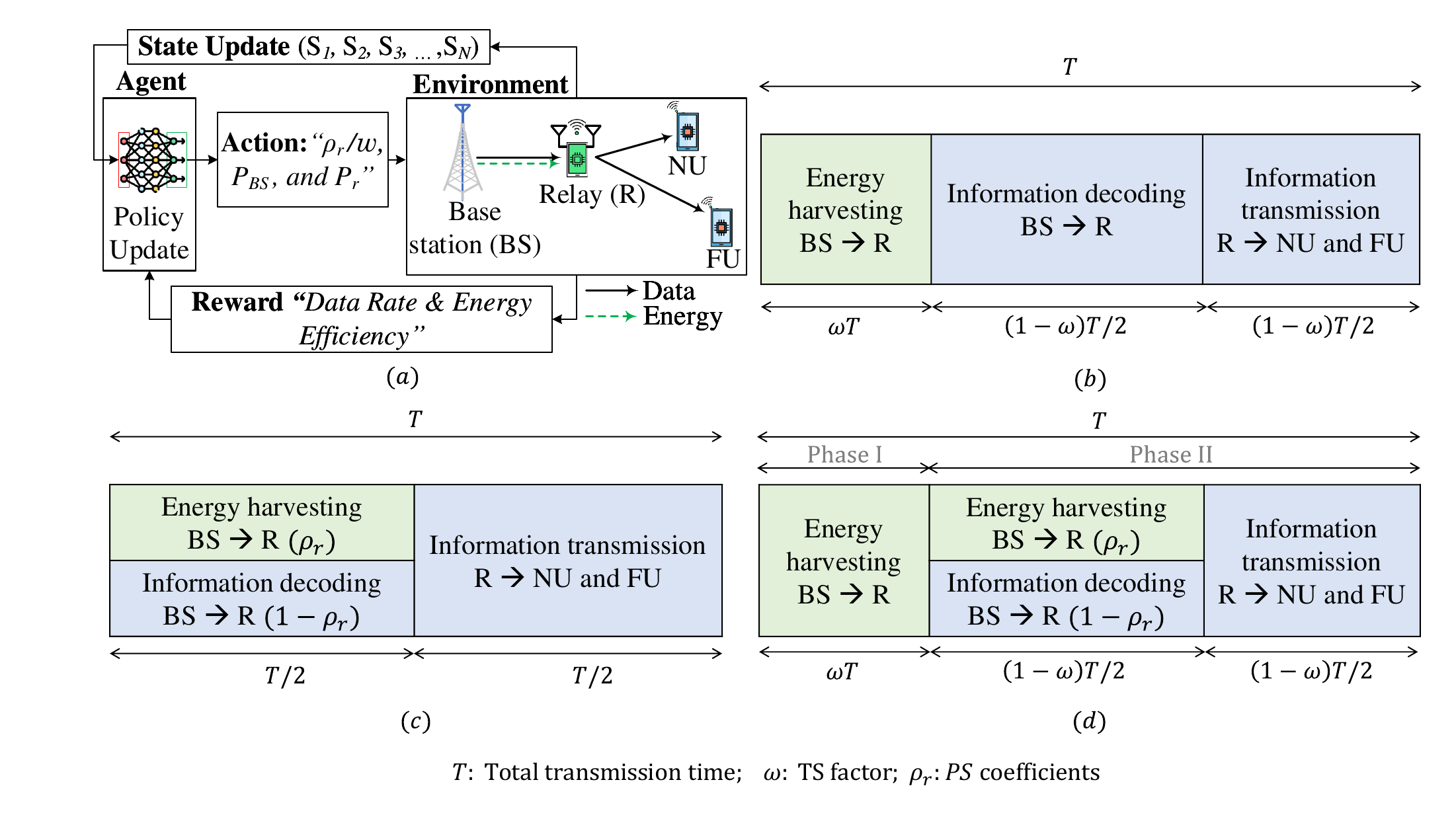}}
\caption{Energy harvesting protocols in C-NOMA (a) System model with DRL optimization approach. (b) TS protocol. (c) PS protocol. (d) HTSPS protocol.}
\label{fig_EH_FD}
\end{figure*}
Enhancing EE in C-NOMA systems requires integrating simultaneous wireless information and power transfer (SWIPT) protocols, where AI can optimize the EH and data transfer trade-offs \cite{drl-ding-2021, drl-shi-2021, dl-shukla-2023}. SWIPT allows relay nodes to harvest energy from RF signals while forwarding information, supporting next-generation connectivity and sustainability. Here, we explore SWIPT approaches within the C-NOMA system model depicted in the environment box in Fig. \ref{fig_EH_FD}(a). The DRL-based system model will be detailed later in this section. Our focus is on three key protocols: time-switching (TS), power splitting (PS), and hybrid TS and PS (HTSPS).

\subsubsection{Time-switching Protocol}In the TS protocol, the relay alternates between EH and data decoding, as shown in Fig. \ref{fig_EH_FD}(b). While simple and effective, it can lead to inefficient time use and reduced throughput, making it less suitable for real-time applications like virtual reality (VR) and augmented reality (AR). AI-driven optimization can help enhance timing decisions and adapt the protocol dynamically to improve performance under varying conditions. 
\subsubsection{Power-splitting Protocol}The PS protocol, illustrated in Fig. \ref{fig_EH_FD}(c), splits the received signal at the relay for EH and information decoding simultaneously. Although this approach enhances real-time processing, optimization is critical for balancing the trade-off between EH efficiency and communication quality, especially in applications like industrial IoT (IIoT) and cognitive city infrastructure. 
\subsubsection{Hybrid Time switching and Power Splitting Protocol}Finally, the HTSPS protocol combines both TS and PS, as depicted in Fig. \ref{fig_EH_FD}(d). It offers flexibility by alternating between the two techniques for better EH and communication performance, making it ideal for advanced applications like autonomous vehicles and remote healthcare. However, the added complexity requires advanced AI algorithms for real-time adaptive management to ensure system reliability and stability in dynamic environments. 

As an advanced machine learning optimization tool, DRL plays a pivotal role in optimizing RF EH and transmission decisions in C-NOMA systems \cite{peng2023energy}. As shown in Fig. \ref{fig_EH_FD}(a), the DRL framework interacts with the system environment to iteratively learn optimal strategies for key decision variables, including the PS ratio $\rho_r$, the TS factor $w$, the BS transmit power $P_{BS}$, and the relay transmit power $P_r$. These parameters are selected based on the active EH protocol, PS, TS, or HTSPS, and dynamically tuned by the agent to maximize data rate and EE. The DRL process is structured into sequential states ($S_1, S_2, ..., S_N$), each representing a unique configuration of system conditions, such as channel gains, residual energy, and interference levels. In each state, the agent takes an action, i.e., adjusting $\rho_r$, $w$, $P_{BS}$, and $P_r$, and receives a reward that reflects the performance impact on throughput and EE. The reward feedback is used to update the agent's policy through a deep Q-networks (DQN) mechanism \cite{liu2022cooperative, liu-2023DQN} or an actor-critic mechanism such as deep deterministic policy gradient (DDPG) \cite{amhaz2024uav}, enabling the agent to make increasingly informed decisions over time. This closed-loop interaction between the agent and the environment ensures continuous learning and adaptation, making the DRL-driven controller capable of handling real-time fluctuations in network dynamics. The DRL approach not only supports protocol-specific parameter tuning but also helps manage the trade-off between EH and information decoding in a unified manner. By learning from experience, the agent can intelligently switch modes (TS, PS, or HTSPS) and reallocate resources to adapt to diverse application needs, from high-throughput scenarios to energy-constrained environments.

Furthermore, future C-NOMA systems can benefit from integrating near-field and far-field wireless power transfer (WPT) techniques, where DRL can orchestrate mode selection and beam control. Near-field WPT ensures high-efficiency charging for proximate devices, while far-field WPT extends energy reach. Leveraging machine learning to jointly optimize both WPT regimes can significantly enhance EH reliability, especially in large-scale deployments like smart cities, autonomous transport, or remote healthcare systems.
\subsection{Cognitive Radio and C-NOMA}
\begin{figure}[!t]
\centering
\centerline{\includegraphics[width=\columnwidth]{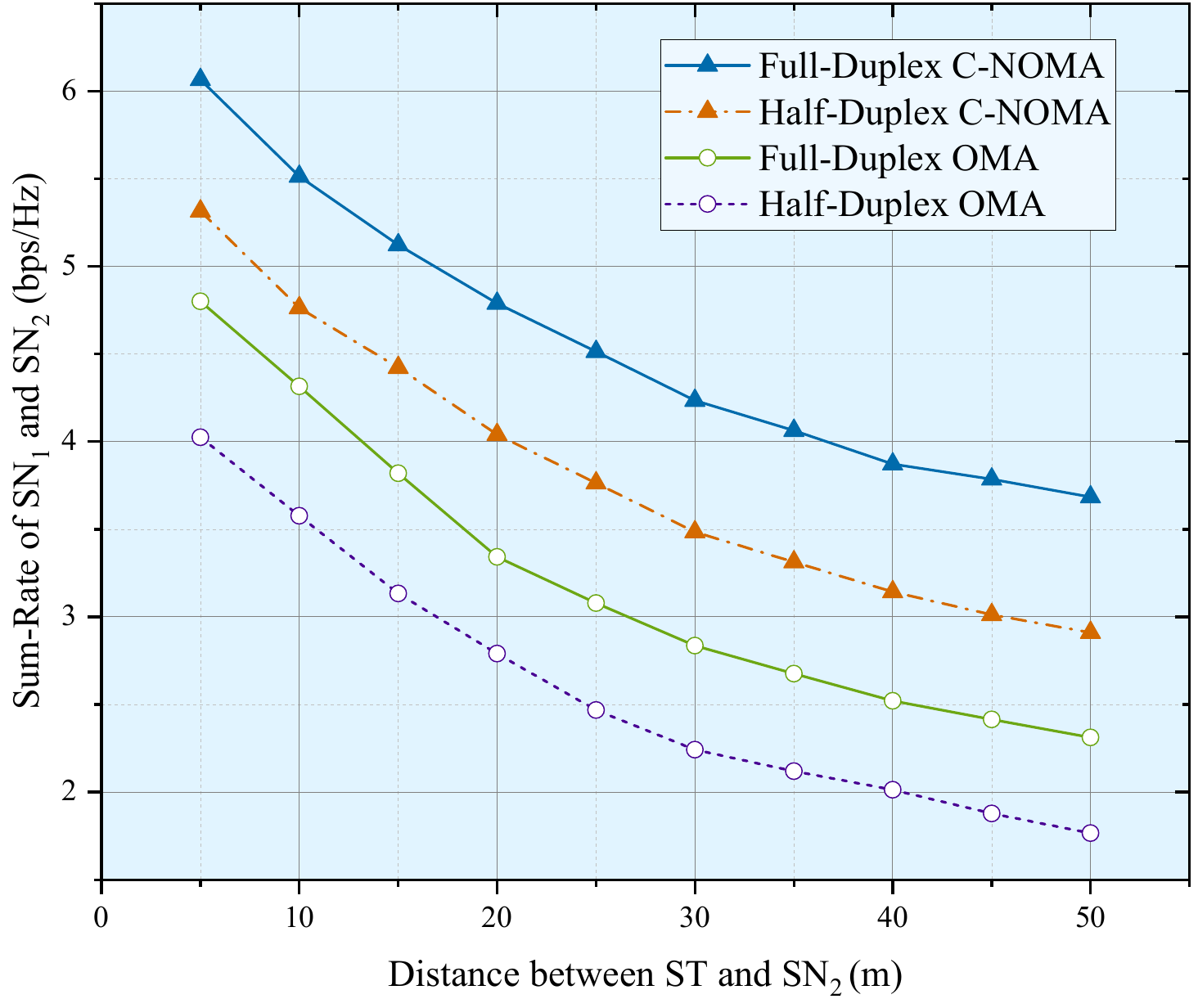}}
\caption{Comparing the performance of C-NOMA and OMA using full-duplex and half-duplex relaying for energy harvesting-based IoT communication scenario.}
\label{fig:Sim_SR}
\end{figure}
Cognitive C-NOMA operates through three primary modes: interweave, underlay, and overlay \cite{zhou-2018,m-Chitra-2023}. While these modes are well-known in the literature, we focus on their unique application within C-NOMA to enhance SE and performance. Leveraging CR technology, these modes optimize spectrum usage, ensure coexistence between primary users (PUs) and secondary users (SUs), and facilitate dynamic spectrum resource management. In the interweave mode, SUs exploit spectrum gaps without interfering with PUs. The underlay mode allows simultaneous transmission between primary and secondary networks, with strict interference limitations to maintain quality-of-service (QoS) of primary communication. Overlay mode superimposes secondary transmissions on primary ones, leveraging primary data knowledge for collaboration and efficient resource use. 

\begin{figure}[!th]
\centering
\centerline{\includegraphics[width=\columnwidth]{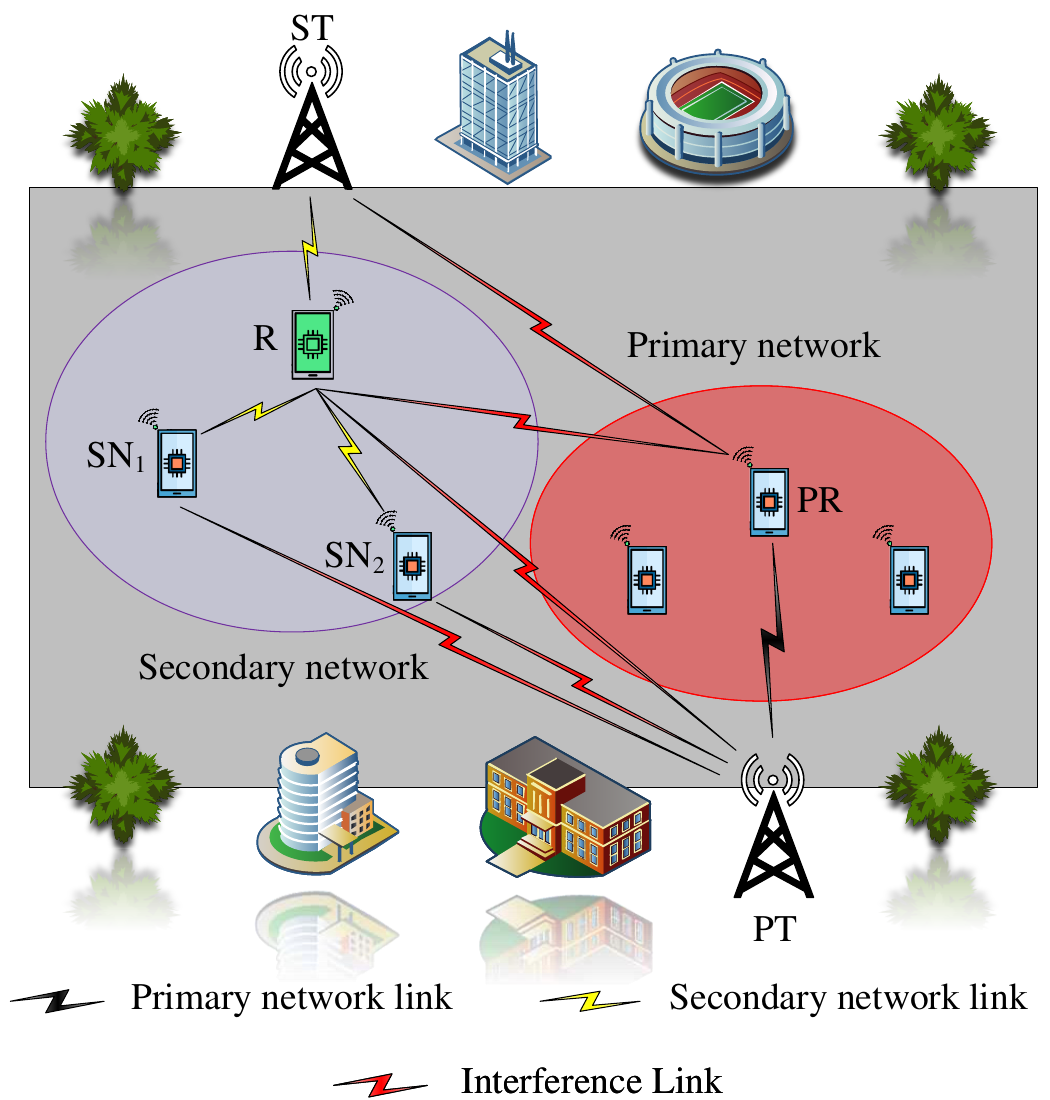}}
\caption{Cognitive C-NOMA in energy harvesting-based IoT communications.}
\label{fig_CR_V2X}
\end{figure}

\begin{figure*}[!th]
\centering
\centerline{\includegraphics[width=1.8\columnwidth]{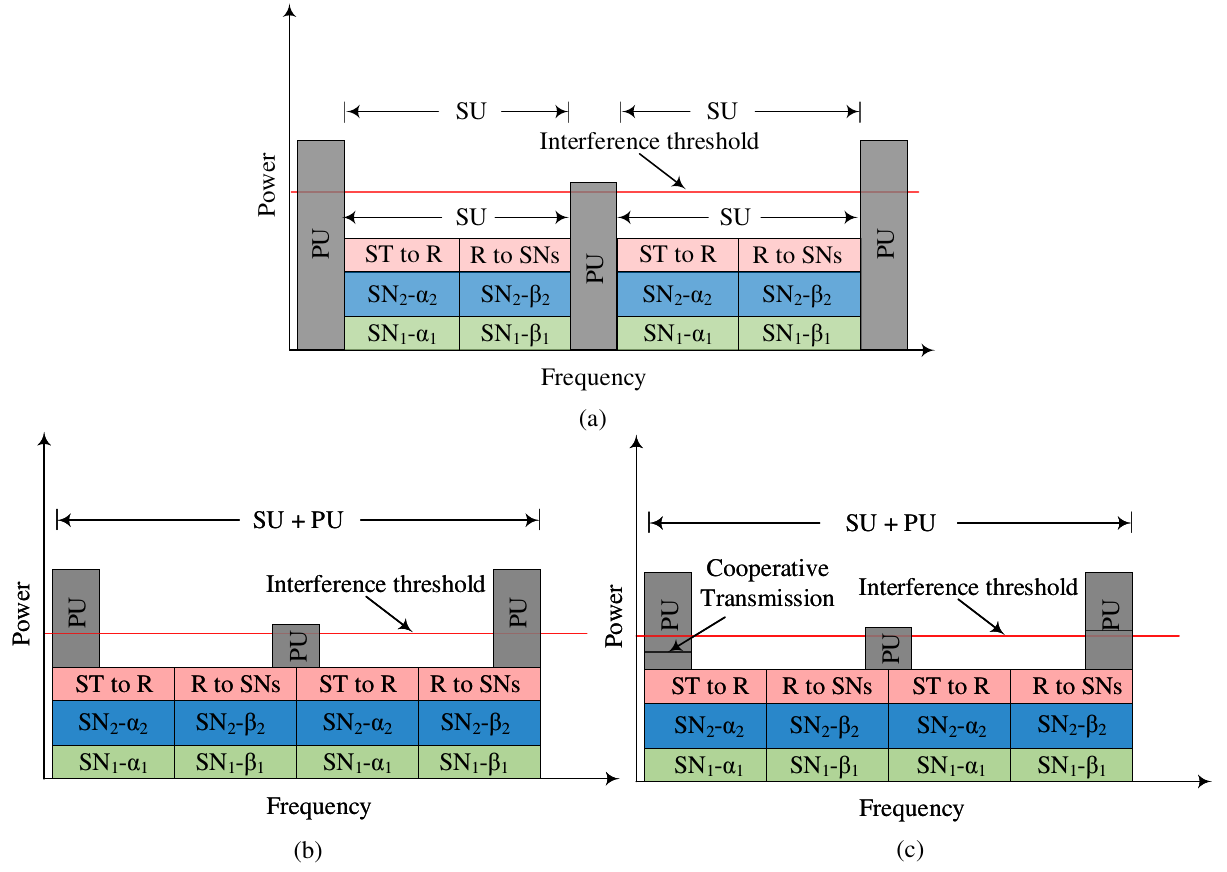}}
\caption{Modes for cognitive C-NOMA in energy harvesting-based IoT communications. (a) Interweave mode. (b) Underlay mode. (c) Overlay mode.}
\label{fig_CR_V2X_Modes}
\end{figure*}

Here, we present our vision for these three cognitive modes and how they operate especially when working with C-NOMA. The cognitive C-NOMA EH IoT communication system model in Fig. \ref{fig_CR_V2X} illustrates the cognitive modes detailed in Figs. \ref{fig_CR_V2X_Modes}(a), (b), and (c). This system features a cooperative NOMA-based secondary network that shares the spectrum with a primary network, consisting of a primary transmitter (PT) and a primary receiver (PR). The secondary network includes a secondary transmitter (ST), a relay (R), and two secondary nodes (SN$_1$ and SN$_2$) representing the NOMA NU and FU, respectively. The ST encodes messages for SN$_1$ and SN$_2$ with power coefficients $\alpha_1$ and $\alpha_2$, transmitting the superimposed message to R, which decodes and re-encodes it with coefficients $\beta_1$ and $\beta_2$. The secondary relay harvests RF energy from the ST via SWIPT technology and renewable sources.

In the interweave mode (Fig. \ref{fig_CR_V2X_Modes}(a)), SU utilizes spectrum gaps without disrupting PU communication. SN$_1$ and SN$_2$ communicate with the ST through the relay during PU inactive periods ensuring efficient spectrum use. The underlay mode (Fig. \ref{fig_CR_V2X_Modes}(b)) allows simultaneous transmission by both networks with a strict interference threshold to maintain the PU's QoS. The secondary nodes as well as the relay transmit within the PU’s allocated spectrum without causing harmful interference. The overlay mode (Fig. \ref{fig_CR_V2X_Modes}(c)) superimposes secondary transmissions on primary ones, enabling SUs to relay data while minimizing interference and maximizing efficiency. Cooperative transmission techniques coordinate secondary and primary nodes to ensure optimal spectrum sharing with an enforced interference threshold protecting primary communications.

Numerical analysis in Fig. \ref{fig:Sim_SR} validates the benefits of C-NOMA and FD in the EH IoT scenario depicted in Fig. \ref{fig_CR_V2X}. The results show the achievable sum-rate of the secondary C-NOMA network as a function of the distance between the ST and SN$_2$, considering optimized power levels below the maximum transmit power of 24 dBm using the particle swarm optimization (PSO) algorithm \cite{salim-2020}. We assume imperfect channel state information (CSI) and imperfect SIC with values of 0.01 and 0.15, respectively. Additionally, for the C-NOMA curves, the PS protocol is applied for RF EH at R, along with renewable energy arrival modeled by a Poisson distribution with an average rate of $3$ mJ/s. As observed, the sum-rate degrades with increasing distance due to path loss effects. However, C-NOMA consistently outperforms conventional OMA techniques regardless of the relaying protocol employed. Furthermore, for the same multiple access technique, FD demonstrates superior performance compared to half-duplex, justifying its selection as the primary relaying scheme in next-generation networks. Notably, the figure illustrates that the energy harvested by the relay supports a higher sum-rate in C-NOMA scenarios, offering a more sustainable approach.

\subsection{Reconfigurable Intelligent Surfaces and C-NOMA}

RISs represent a transformative technology capable of manipulating the wireless propagation environment by intelligently adjusting the amplitude, phase, or polarization of incident electromagnetic waves \cite{hu2020reconfigurable,kazim2022application}.
Through careful reconfiguration, RISs enable significant gains in SE, EE, and reliability while maintaining a low-cost and nearly passive architecture.
For instance, RISs can be integrated with energy harvesting circuits \cite{sharma2025spectrum,zhao2023ris} to sustain UAV-mounted relays, ensuring prolonged operation and stable support for C-NOMA users, especially in infrastructure-scarce scenarios. Moreover, accurate channel modeling for RIS-aided networks has become crucial to characterize the unique cascaded propagation effects introduced by the surface, including near-field and far-field behaviors \cite{zhao2023ris,huang2023integrating}.
These characteristics position RISs as highly attractive complements to C-NOMA systems, where improving user fairness, enhancing SIC, and optimizing spectrum utilization are of paramount importance \cite{chauhan-2022-SIC,tusha-2024Int}. Here, we focus on the RISs' deployment strategies and operational modes, which play a distinct role in shaping the future of C-NOMA.
\subsubsection{RIS Deployment Strategies for C-NOMA}

The deployment architecture of RISs has a direct impact on their ability to support C-NOMA networks. 
In traditional deployments, RISs are installed on fixed infrastructure such as building facades, poles, or towers \cite{zhou2024secure}. Fixed RISs serve to enhance the communication environment by overcoming non-line-of-sight limitations, strengthening the effective channel gains, and improving inter-user fairness, which is vital for the efficient operation of SIC in C-NOMA clusters.
For example, Fig.~\ref{fig_Big_Pict} illustrates a terrestrial setup where RIS-equipped buildings improve the connectivity between a BS and vehicular users, facilitating V2X-based C-NOMA operations and supporting applications such as traffic flow optimization and hazard prediction. 
Despite their benefits, fixed RISs are constrained by their static nature, rendering them less effective in rapidly changing environments or dynamic user distributions.

To address these limitations, UAV-mounted RISs have been proposed as a promising alternative \cite{RIS-10816259, saif2024improving}. By mounting RIS panels on UAVs, it becomes possible to dynamically reposition the RISs according to user mobility, traffic load, or environmental factors.
In a C-NOMA context, UAV-mounted RISs can optimize the BS-to-user links by adjusting both their trajectory and their phase shift configurations \cite{zhang2022capacity}, thereby enhancing signal strength to distant users and balancing the SINR disparity needed for effective SIC decoding.
However, UAV-mounted RISs introduce new challenges, including limited onboard energy, real-time control under mobility, and flight stability constraints \cite{sahoo2022toward}.
DRL algorithms have been suggested to jointly optimize UAV trajectories, RIS phase shifts, and energy harvesting strategies to extend UAV endurance and network performance \cite{drl-guo-2023,kokare2025reinforcement}.

\begin{figure}[t]
\centering
\centerline{\includegraphics[width=\columnwidth]{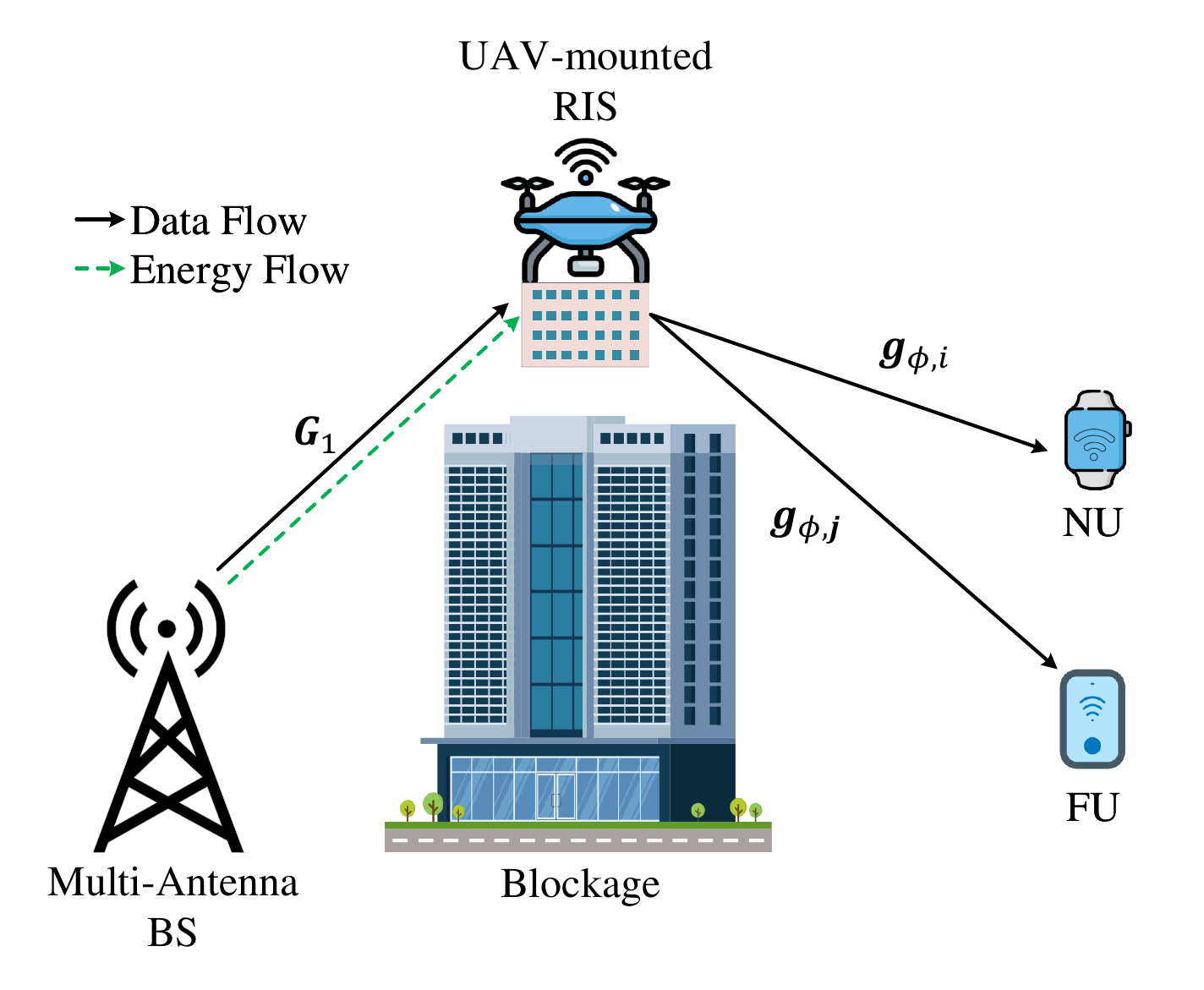}}
\caption{UAV-mounted RIS with RF energy harvesting capabilities for RIS elements relaying C-NOMA IoT nodes.}
\label{fig_UAV-mountedRIS}
\end{figure}

The architecture illustrated in Fig.~\ref{fig_UAV-mountedRIS} highlights a UAV-mounted RIS-enabled C-NOMA system integrated with RF EH. Here, a UAV-mounted RIS acts as a smart reflector between a multi-antenna BS and two distinct user types, namely a NU and a FU. The BS transmits a power-domain multiplexed signal using superposition coding to simultaneously serve both users, while the RIS adaptively configures its phase shifts to assist particularly the FU, which experiences harsher propagation conditions. The system operates over a cascaded MIMO channel composed of the BS-to-RIS link and the RIS-to-user links. Let $\mathbf{G}_{1} \in \mathbb{C}^{N \times M}$ denote the MIMO channel between the BS and the RIS, where $N$ is the number of RIS elements and $M$ is the number of BS antennas.
The effective cascaded channel between the BS and a user $u \in \{i,j\}$ through the RIS is modeled as
\begin{equation}
    \mathbf{h}_{\text{BS-}u} = \mathbf{g}_{\phi,u} \mathbf{\Phi} \mathbf{G}_{1},
\end{equation}
where $\mathbf{g}_{\phi,u} \in \mathbb{C}^{1 \times N}$ represents the RIS-to-user channel, and $\mathbf{\Phi} = \text{diag}(e^{j\theta_1}, \dots, e^{j\theta_N})$ denotes the RIS phase shift matrix. The transmitted signal from the BS is expressed as
\begin{equation}
    x = \sqrt{P_i}s_i + \sqrt{P_j}s_j,
\end{equation}
where $s_i$ and $s_j$ are the information symbols intended for the NU and FU, respectively, and $P_i < P_j$ ensures that SIC at the NU is feasible.

To enhance the endurance of the UAV-mounted RIS, a TS based EH model is employed \cite{liu2017qos}. Specifically, each transmission time block $T$ is partitioned into two phases: during the EH phase of duration $\tau T$, all RIS elements harvest energy from incident RF signals; during the transmission phase of duration $(1-\tau)T$, the RIS performs intelligent reflection to assist the BS-user links. Assuming a linear EH model, the harvested energy at the RIS during time slot $t$ can be expressed as
\begin{equation}
    E_{\text{RIS}}(t) = \tau(t) \eta \sum_{n=1}^{N} \| \mathbf{g}_{n}^H \mathbf{X}(t) \|^2,
\end{equation}
where $\mathbf{g}_{n}$ represents the channel vector from the BS to the $n$-th RIS element, $\mathbf{X}(t)$ is the transmitted signal matrix at time $t$, and $\eta$ is the RF energy conversion efficiency \cite{peng2023energy}.

To further enhance UAV sustainability, an advanced hybrid-domain harvesting strategy can be considered. This approach allows a subset of RIS elements to reflect signals while the remaining elements continue harvesting energy even during the transmission phase, significantly boosting EE without sacrificing communication performance \cite{peng2023energy,salim2025energy,salim2025energyharvesting}. The achievable rate in our case for each user, considering the active transmission phase, is given by
\begin{equation}
    R_u = (1-\tau)\log_2(1+\gamma_u),
\end{equation}
where $\gamma_u$ denotes the instantaneous SINR
\begin{align}
    \gamma_j &= \frac{|\mathbf{h}_{\text{BS-}j} \mathbf{w}_2|^2}{|\mathbf{h}_{\text{BS-}j} \mathbf{w}_1|^2 + \sigma^2}, \\
    \gamma_i &= \frac{|\mathbf{h}_{\text{BS-}i} \mathbf{w}_1|^2}{\sigma^2},
\end{align}
with $\mathbf{w}_1$ and $\mathbf{w}_2$ representing the BS beamforming vectors, and $\sigma^2$ denoting the noise variance. Joint optimization of the TS ratio $\tau$, the RIS phase shift matrix $\mathbf{\Phi}$, beamforming vectors $\{\mathbf{w}_1, \mathbf{w}_2\}$, and power allocations $\{P_i, P_j\}$ is essential to maximize the overall C-NOMA throughput while ensuring sustainable operation of the UAV-mounted RIS system.

Another key metric is EE, defined as the total achievable rate per unit of consumed power. A representative EE expression is: \begin{equation} \text{EE} = \frac{\sum_{u \in {i,j}} R_u}{P_{\text{BS}} + P_{\text{RIS}} + P_{\text{UAV}} + P_{\text{c}}}, \end{equation} where $P_{\text{BS}}$, $P_{\text{RIS}}$, and $P_{\text{c}}$ denote the BS transmit power, RIS element power, and circuit power, respectively, while $P_{\text{UAV}}$ represents the UAV’s propulsion power, which depends on its speed, altitude, and hover time \cite{batteryweight-yan-2021, pmsg-li-2023}. While rate and EE are widely optimized, other objectives, such as EH efficiency, user fairness, secrecy rate, and latency, are increasingly considered to address broader design trade-offs in UAV-assisted C-NOMA systems.

\subsubsection{RIS Operational Modes for C-NOMA}

\begin{figure*}[!t]
  \centering
  \includegraphics[width=0.95\linewidth]{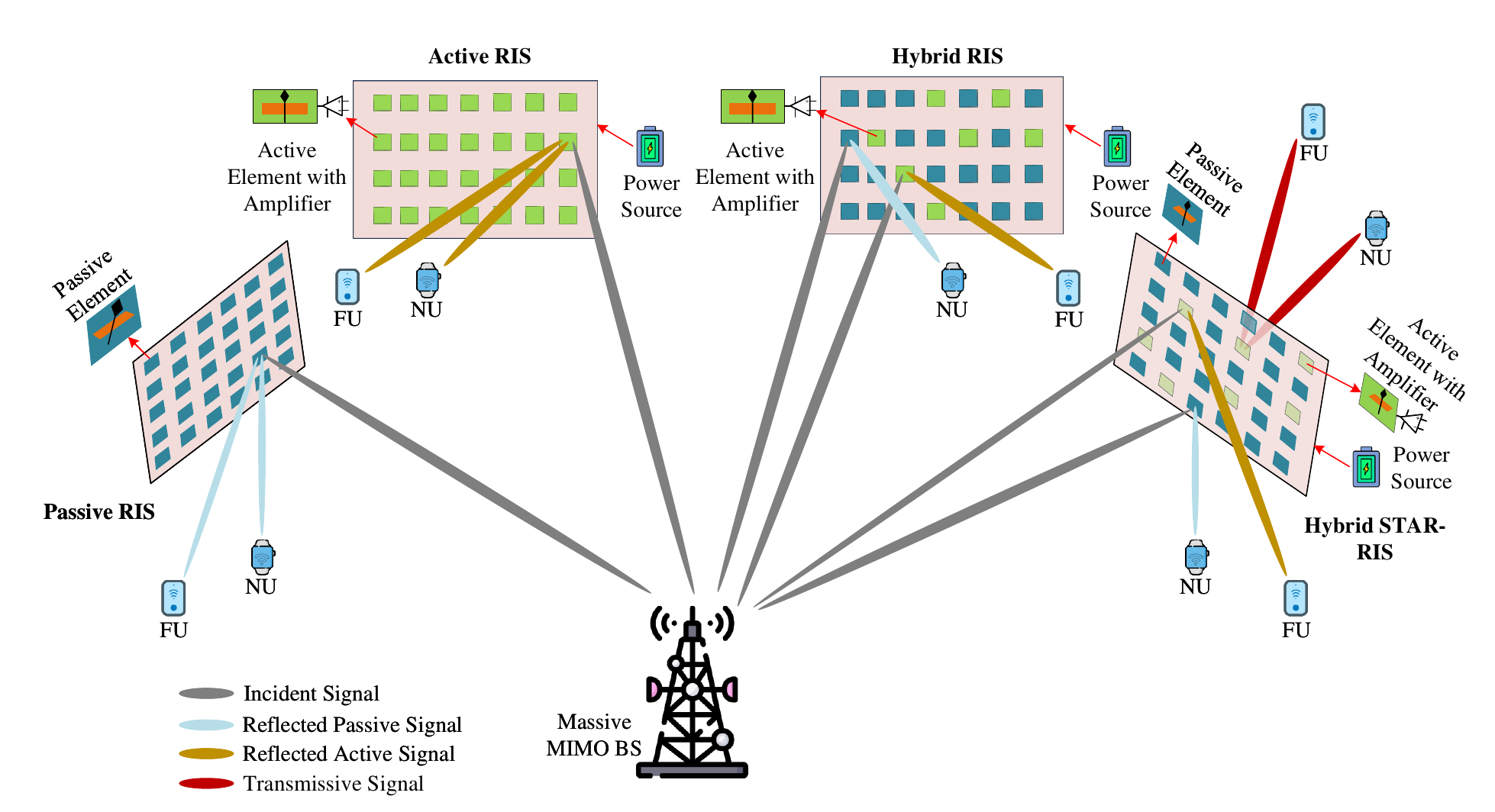}
  \caption{RIS operational modes in C-NOMA.}
  \label{fig:RIS_Op}
\end{figure*}

RISs can operate in different modes depending on their architecture and hardware capabilities. In C-NOMA systems, where NUs and FUs coexist under power-domain multiplexing, the choice of RIS mode significantly affects link quality, SIC reliability, and overall system efficiency. Fig.~\ref{fig:RIS_Op} illustrates the operation of these RIS types and their impact on C-NOMA user targeting, where a detailed description is presented as follows.

\textbf{1) Passive RIS:} 
Consisting only of phase-shifting elements without amplification, passive RISs are highly energy-efficient and suitable for large-scale, low-cost deployments \cite{secure-ris-lv-2022,saikia2024ris}. They are often used to assist NUs, where the SNR is already adequate and phase alignment enhances decoding. However, for FUs, their effectiveness can be limited due to double path loss and the lack of amplification, especially when the RIS is not optimally placed. That said, with proper deployment and intelligent beamforming, passive RISs can still significantly benefit FUs, particularly in line-of-sight or low-interference conditions, though possibly requiring larger element arrays to maintain user fairness and SIC reliability.

\textbf{2) Active RIS:} 
Active RISs integrate active reflection amplifiers within their reflecting elements to boost weak incident signals, thereby significantly enhancing the received SINR, particularly for FUs in C-NOMA systems \cite{active-ahmed-2024}. This capability makes them especially valuable in scenarios where direct or reflected links to FUs are severely attenuated, improving decoding reliability and SIC performance. However, the amplification comes at the cost of increased power consumption and hardware complexity. Moreover, the introduced amplification noise must be carefully controlled to prevent degradation in overall system performance, necessitating advanced signal processing and power management strategies.

\textbf{3) Hybrid RIS:}
Hybrid RISs integrate both active and passive reflecting elements, providing a versatile architecture capable of simultaneously supporting NUs and FUs \cite{tishchenko-2025}. In typical configurations, active elements are oriented toward FUs to compensate for severe signal attenuation, while passive elements assist NUs with energy-efficient phase adjustments. This hybrid design strikes a balance between performance and power consumption, enhancing the reliability of SIC while containing hardware complexity and energy overhead.

\textbf{4) Hybrid Simultaneously Transmitting and Reflecting RISs (STAR-RISs):} 
Hybrid STAR-RISs extend conventional RIS functionality by integrating both active and passive components across a surface that enables dual-sided coverage \cite{ahmed-2023Survey, magbool2024survey}. This advanced architecture can support multiple C-NOMA user pairs, for instance, a NU-FU pair on the reflecting side and another on the transmitting side, by adaptively splitting the incident signal into reflected and transmitted components. Active elements with amplification can be allocated toward weaker FUs, while passive elements manage near-user assistance or energy-efficient links. Such a configuration greatly enhances flexibility, spatial multiplexing, and fairness, especially in full-space deployments like UAV-mounted or building-embedded RISs \cite{zhu2023optimized}. However, hybrid STAR-RISs require separate optimization of reflection, transmission, and gain coefficients across user clusters, significantly increasing design complexity, synchronization overhead, and control signaling demands.

\vspace{2mm}
\vspace{2mm}

\begin{table}[!t]
\caption{Performance Impact of RIS Operational Modes in C-NOMA}
\centering
\scalebox{0.73}{
\begin{tabular}{|l|c|c|c|c|}
\hline
\textbf{Feature} & \textbf{Passive RIS} & \textbf{Active RIS} & \textbf{Hybrid RIS} & \textbf{STAR-RIS} \\
\hline
Power Consumption & Very Low & High & Moderate & Moderate \\
\hline
Amplification & \xmark & \cmark & Partial & Partial \\
\hline
User Targeting & NU & FU & NU + FU & NU + FU \\
\hline
SIC Reliability & Limited (for FU) & High & Improved & Improved \\
\hline
SE & Moderate & High (for FU) & Balanced & High \\
\hline
EE & Excellent & Low & Balanced & Moderate \\
\hline
Complexity & Low & High & Moderate & High \\
\hline
Full-Space Coverage & \xmark & \xmark & \xmark & \cmark \\
\hline
Typical Application & NU boosting & FU compensation & Joint enhancement & 360$^\circ$ \\
\hline
\end{tabular}}
\label{tab:ris_modes}
\end{table}

To summarize, passive RISs improve decoding for NUs but are ineffective for FUs, reducing fairness. Active RISs significantly boost FUs' SINR and enable robust SIC but at a power and noise cost. Hybrid RISs offer a practical compromise, selectively boosting FUs while maintaining EE for NUs. Hybrid STAR-RISs provide the most flexible user support and highest SE potential, particularly for multi-cluster full-space C-NOMA, but demand the most complex optimization. Whether fixed or UAV-mounted, RISs enhance C-NOMA by expanding coverage and enabling adaptive resource control. This synergy is further amplified by AI-driven optimization, which adjusts RIS configurations based on CSI, mobility, and user density \cite{hashi2022channel,zhou2024secure}. Such intelligence is essential for managing the trade-off between EE and communication reliability, particularly when switching between passive, active, or hybrid modes. A comparison of these RIS types in terms of their impact on C-NOMA performance is summarized in Table~\ref{tab:ris_modes}.

\subsection{Space-Air-Ground Integrated Networks and C-NOMA}
SAGINs are a key enabler for scaling the capabilities of C-NOMA in future 6G systems that demand ubiquitous coverage, massive connectivity, and low-latency communications \cite{ai-sagin-bakambekova-2024}. Comprising three interlinked layers: space-based (e.g., low earth orbit (LEO) satellites), air-based (e.g., UAVs and high-altitude platforms), and ground-based (e.g., BSs and IoT devices), SAGINs leverage the complementary strengths of each domain to support a broad spectrum of use cases. Satellites provide global reach and high-capacity backhaul, UAVs offer flexible and adaptive relaying, and ground infrastructure ensures local processing and high data rates.

By integrating C-NOMA into this hierarchical framework, SAGINs can serve ultra-dense IoT deployments, real-time V2X communications, and remote monitoring in both urban and underserved regions. C-NOMA’s cooperative strategies, such as user relaying and cognitive spectrum access, extend naturally across layers, improving signal propagation, user decoding, and spectral and EE. For example, as depicted in Fig.~\ref{fig_Big_Pict}, vehicles in a V2X scenario may retrieve satellite-based weather updates while UAVs facilitate low-latency message relaying between nearby nodes.
The incorporation of RISs into SAGINs further elevates the adaptability and performance of C-NOMA systems \cite{ris-liu-2024, envisioning-xu-2021}. By enabling programmable reflection, RISs support beam steering, interference suppression, and path compensation across dynamic environments. Whether deployed on satellites, UAVs, or mobile vehicles, RISs enhance multi-user communication quality, boost SINR, and enable functionalities such as EH and semantic communication. However, to fully exploit RIS-aided C-NOMA in SAGINs, joint beamforming optimization across mobile and passive components must account for channel mobility, latency, and imperfect CSI \cite{yu2024beamforming,lu2020robust,kong2021beamforming}. Large-scale deployments introduce challenges such as signal alignment, multi-layer coordination, and synchronization overhead. In constrained scenarios like V2V backscatter communication, RISs offer energy-efficient enhancements by reusing satellite signals for low-power links. To address these complexities, machine learning and DRL are increasingly critical for real-time control and adaptive configuration in RIS-enabled SAGINs \cite{sobhi-givi-2025}.

In summary, integrating C-NOMA with RIS-empowered SAGINs paves the way for resilient, scalable, and energy-efficient 6G networks. This synergistic architecture is well-suited for mission-critical and data-intensive applications, such as autonomous transport, environmental monitoring, and emergency communications, particularly in remote or infrastructure-limited regions. As the convergence of these technologies matures, it holds the potential to redefine connectivity in dynamic, heterogeneous network environments.

\subsection{ISAC-Assisted Semantic Communication in C-NOMA: Framework and Edge-Driven UAV Scenario}
Semantic communication marks a paradigm shift in wireless systems, evolving from conventional bit-level data transmission to goal-oriented message exchange \cite{yang2022semantic,10038657,chaccour-2024}. Instead of transmitting raw or symbol-level information, semantic communication enables devices to extract and convey only task-relevant content \cite{da-costa-2023}. This aligns naturally with the philosophy of C-NOMA, where spectrum sharing among multiple users must be managed efficiently in interference-limited environments. By prioritizing meaning over quantity, semantic communication reduces transmission overhead, enhances SE, and improves scalability in ultra-dense deployments \cite{adaptive-yan-2024}. Deep learning techniques further enhance this capability by enabling neural encoders and decoders to extract, compress, and reconstruct high-level semantic features, ensuring robustness even under imperfect CSI \cite{interference-zhang-2024,cooperative-kaewpuang-2023,gan-mao-2024,flag-huang-2024}.

Parallel to this, ISAC has emerged as a complementary paradigm that uses shared radio resources for both communication and sensing functions \cite{liu2022integrated,semantic-zhu-2024}. In C-NOMA contexts, ISAC provides critical environmental awareness while maintaining efficient connectivity. A promising vision is that of semantic-aware ISAC, where raw sensed data are interpreted through a task-oriented lens, and only compressed, relevant semantic features (e.g., "obstacle ahead", "traffic congestion", or "low visibility") are transmitted. This evolution enhances both decision latency and system EE, making it ideal for next-generation intelligent systems \cite{semantic-shao-2024,sharshir2025using}.

\begin{figure}[!h]
\centering
\centerline{\includegraphics[width=\columnwidth]{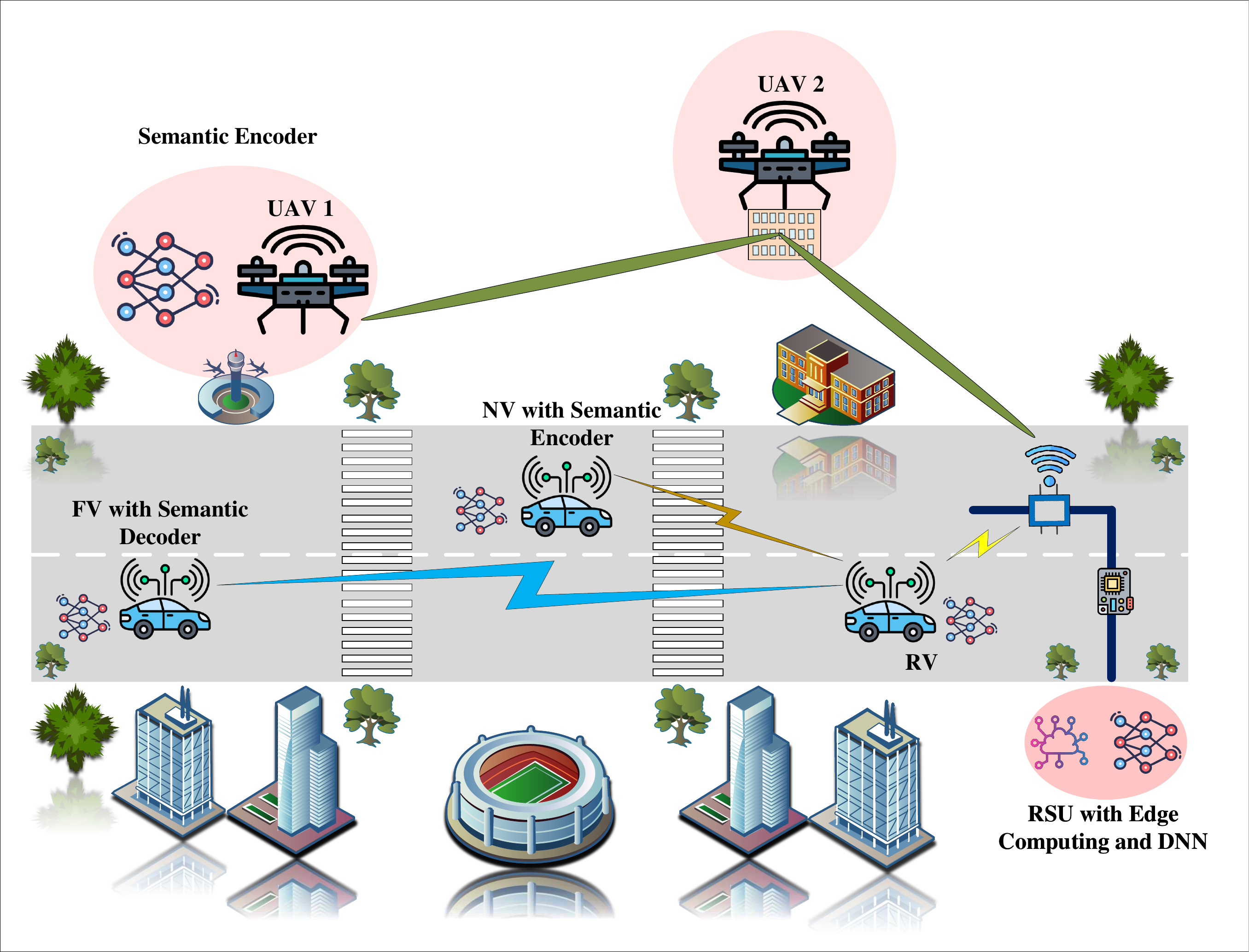}}
\caption{Integration of C-NOMA with ISAC, edge computing, and RIS in vehicular networks.}
\label{fig_SemCom_ISAC}
\end{figure}

A practical scenario of this vision is illustrated in the vehicular scenario shown in Fig.~\ref{fig_SemCom_ISAC}. In this case, C-NOMA is integrated with ISAC, edge computing, and RIS-aided UAV communications to support intelligent vehicular networks. UAV 1 captures traffic scenes and performs real-time semantic encoding using a DNN. The extracted features are relayed to a roadside unit (RSU) via a RIS-supported UAV 2. The RSU, equipped with edge computing and the same DNN architecture, performs model compression (e.g., pruning and quantization \cite{9398534,EDROPOUT2022,THINEt2022,STNET_2022,Pruning24,fei-2022,wang-2023Q,9398534,shi-2024Q}) and generates lightweight i2nference models suitable for resource-constrained vehicles. These models and semantic features are then broadcast to near and far vehicles (NV and FV) in a power-domain superposition format using a relay vehicle (RV) based on C-NOMA principles.

This framework offloads computational burden from mobile devices, ensures robust feature transmission, and enhances spectrum reuse. RIS improves the received signal quality and coverage between UAVs, RSUs, and vehicles. However, high-mobility environments introduce Doppler and delay spread distortions, requiring adaptive beamforming, robust channel tracking, and AI-enhanced feature recovery to preserve decoding accuracy \cite{9919739}. These challenges can be mitigated through end-to-end learning and joint channel-feature co-optimization strategies. The resulting system supports reliable, low-latency communication and perception fusion in real-time, which is essential for autonomous driving, cooperative perception, and traffic optimization. The convergence of C-NOMA, semantic communication, ISAC, and RIS-aided UAVs establishes a unified, energy-efficient, and context-aware framework for cognitive transportation systems and other 6G-era applications.
\subsection{Orchestration Among Envisioned Applications} 

Figure~\ref{fig_Big_Pict} visualizes the envisioned convergence of C-NOMA with complementary technologies across diverse application domains in next-generation networks. This section offers a deeper look into how these integrations enable adaptive, intelligent, and energy-efficient communications in scenarios characterized by mobility, ultra-low latency, and massive connectivity. We assume that all relay nodes operate in FD mode, aligning with emerging 6G trends. FD relaying, when combined with C-NOMA and advanced SIC, significantly enhances SE and minimizes end-to-end delay, which is crucial for mission-critical IoT applications.

\textbf{1) Remote Healthcare and Emergency Response:}
In real-time healthcare and disaster relief scenarios \cite{abdalzaher2024enhancing,en16010495}, UAV-mounted RIS relays combined with FD-C-NOMA offer reliable connectivity under uncertain and time-sensitive conditions. For instance, UAVs equipped with sensors and RIS panels can gather patient vitals or accident-site data and reflect it toward mobile medical units or hospitals. The FD-C-NOMA architecture allows simultaneous transmission to multiple units using power-domain multiplexing, while RIS phase control enhances link quality. Semantic communication with DNNs allows transmission of only high-level features (e.g., "cardiac anomaly detected") instead of raw sensor data, reducing bandwidth consumption and latency. This ISAC-assisted setup supports ultra-reliable low-latency communications and resilience under high user mobility or network congestion.

\textbf{2) Autonomous Transport and Smart Cities:}
In V2X systems, C-NOMA’s spectrum-sharing flexibility enables dense user environments such as traffic intersections, where multiple cars and drones share the same frequency band. Drones as PUs performing real-time sensing can relay traffic congestion or weather conditions to a BS via RIS nodes, while C-NOMA secondary networks (e.g., RSUs or buses) exploit underutilized spectrum. Joint UAV-RIS trajectory optimization and spectrum access can be managed via DRL-based controllers, dynamically adapting to traffic density and urban topology. Such configurations fulfill latency and reliability constraints, essential for cooperative driving and hazard avoidance.

\textbf{3) Industrial IoT and Robotic Systems:}
C-NOMA facilitates large-scale device access in factory automation where hundreds of sensors and robotic units operate concurrently. Hybrid RIS configurations (with active elements directed toward far edge devices) ensure uniform SINR distribution across the factory floor, boosting decoding success for both near and FUs. In robotic control, haptic feedback and real-time updates require extremely low delay and jitter. Here, FD-C-NOMA enables simultaneous control and feedback loops, while semantic ISAC offloads redundant information, ensuring efficient and reliable machine communication under tight QoS constraints.

\textbf{4) XR and Digital Twins in Sports and Training:}
In immersive sports applications, a coach can control football robots via haptic devices, with digital twin feedback rendered in real-time. C-NOMA supports simultaneous command dissemination and sensory feedback using power-domain multiplexing, ensuring precise and timely actuation. UAV-mounted RISs provide dynamic coverage across a wide field, while EH modules ensure prolonged UAV operation. Semantic encoding compresses motion or event descriptors, enabling low-latency closed-loop control under constrained bandwidth.

\textbf{5) Cross-Cutting Consideration – Energy Harvesting:}
Across all domains, EH from RF and renewable sources enhances sustainability, especially for battery-limited UAVs and distributed RIS elements. Integrating EH-aware resource management into the C-NOMA protocol stack is critical to maintaining long-term operability in dynamic environments. Optimization objectives extend beyond rate and EE to include EH efficiency, fairness, latency, and even security.

In summary, the orchestration of C-NOMA with RISs, UAVs, semantic communication, and ISAC enables a unified, context-aware communication fabric. By tailoring power, location, spectrum, and data semantics jointly, these technologies collectively address the multifaceted demands of next-generation intelligent systems.

\section{Literature on C-NOMA and Emerging Technologies} \label{Sec_RW}
\subsection{RF Energy Harvesting and C-NOMA Integration}
The integration of EH into C-NOMA systems has gained significant attention due to its potential to ensure sustainable connectivity in power-constrained wireless environments. Various protocols, including TS, PS, hybrid SWIPT, and more advanced EH strategies, have been proposed in diverse scenarios such as UAV networks, IoT systems, and RIS-assisted networks. Initial efforts explored adaptive cooperative strategies where strong users harvest energy and relay messages for weak users, ensuring fairness in C-NOMA networks. For example, \cite{li-2020Adaptive} proposed an adaptive cooperative NOMA framework utilizing a FD user-relay powered via EH to guarantee reliability for the weak user. In another work, \cite{mittal2023performance} analyzed a NOMA network with non-linear EH at the relay under Nakagami-m fading, providing outage expressions for user performance. Multiple-input single-output (MISO)-based C-NOMA systems with EH were studied in \cite{karthik2021cooperative}, which demonstrated capacity and bit error rate (BER) gains over conventional NOMA using interference cancellation. The impact of jamming and interference on secrecy performance with RF EH in cooperative NOMA was addressed in \cite{thakur2021novel}, revealing optimal jamming conditions for secure EH. TS relaying in C-NOMA with an EH DF relay was considered in \cite{sarkar2022outage} and \cite{ghosh2022energy}, both deriving closed-form outage probability (OP) expressions and confirming the existence of optimal TS coefficients. In \cite{2024-rate-salim}, a practical cognitive C-NOMA framework with hybrid-powered FD relays was proposed, addressing sum-rate and EE tradeoffs under realistic impairments using a PS-based RF/renewable EH model and a novel relay selection algorithm.

SWIPT-based systems have also received attention. \cite{neole2024dynamic} presented a hybrid TS/PS framework for SWIPT in C-NOMA, showcasing significant improvements in OP and achievable rate. Similarly, \cite{liu2022joint} and \cite{ren-2022RIS} applied SWIPT with RIS-assisted cooperative NOMA networks, optimizing PS ratios and RIS phase shifts to improve cell-edge throughput. A hybrid TS-PS protocol was further analyzed in \cite{zhang-2022Hybrid} to demonstrate enhanced diversity gains via transmit antenna selection. EH from UAV sources is another focal area. Works such as \cite{bajpai-2025Wi, shaikh2024energy, hu2021physical} proposed UAV-enabled cooperative user-to-user (U2U) and cellular links powered via SWIPT and WPT. In particular, \cite{bajpai-2025Wi} studied FD cooperative U2U links with fluctuating two-ray fading, and \cite{hu2021physical} incorporated cooperative jamming to support physical-layer security. For underwater acoustic systems, \cite{mittal2023energy} analyzed EH at a UAV-based relay using PS to power DF transmission toward underwater sensor nodes.

RISs also play a critical role in enhancing EH-based C-NOMA. The work in \cite{dey2024partitioning} introduced mode-switching C-NOMA with an EH-enabled NU and RIS partitioning. Similarly, \cite{liu2022joint, zhang-2022Hybrid, vu-2023STAR} studied SWIPT-enabled RIS-NOMA systems, showing the trade-offs between outage performance and EE. Several other studies addressed joint resource optimization for energy-efficient EH-CNOMA networks. For instance, \cite{kumar-2023} investigated deep learning-based power control for UAV-based spectrum sharing with EH, while \cite{jalali-2024J} proposed a terahertz-band SWIPT NOMA system for miniature UAVs. The authors in \cite{nguyen-2023Se} considered UAV-enabled mobile edge computing (MEC) with WPT and secrecy constraints, demonstrating the importance of trajectory optimization. Additionally, the work in \cite{shen-2020} evaluated UAV-assisted wireless powered sensor networks using cooperative NOMA relaying, optimizing pairing and altitude for EH. These contributions collectively demonstrate that integrating EH into C-NOMA provides substantial benefits in terms of energy sustainability, SE, and reliability. Furthermore, the choice of EH protocol, either TS, PS, hybrid SWIPT, or RF-based, significantly influences system performance and optimization complexity, particularly in dynamic or mobility-aided scenarios such as UAV or RIS-assisted networks. Table \ref{tab:eh_cnoma_lit} highlights the summary of related work on EH in C-NOMA systems.

\begin{table*}[ht]
\centering
\caption{Summary of Related Work on Energy Harvesting in C-NOMA Systems}
\begin{tabular}{|l|p{5.0cm}|p{1.7cm}|p{2.5cm}|p{5.5cm}|}
\hline
\textbf{Ref.} & \textbf{Scenario / Focus} & \textbf{EH Protocol} & \textbf{Metric(s)} & \textbf{Key Insight} \\
\hline
\cite{li-2020Adaptive} & Adaptive cooperative NOMA with user-relaying & Adaptive RF PS & OP, reliability & FD user relays weak user’s data using harvested energy \\
\hline
\cite{mittal2023performance} & Relay-assisted NOMA with non-linear EH & Non-linear RF PS & OP & Non-linear EH improves relay reliability in fading \\
\hline
\cite{karthik2021cooperative} & Cooperative MISO-NOMA & RF PS & BER, capacity, SE & EH enhances MISO-NOMA performance vs. baseline \\
\hline
\cite{thakur2021novel} & Interference-aided EH in secure NOMA & RF TS & Secrecy outage, ergodic secrecy rate & Interference improves secure EH in C-NOMA \\
\hline
\cite{sarkar2022outage, ghosh2022energy} & EH DF relay in NOMA & RF TS & OP & TS achieves optimal relay performance \\
\hline
\cite{neole2024dynamic} & Dynamic power allocation in EH C-NOMA & Hybrid RF TS/PS & Outage, rate, BER & TS/PS tradeoff improves rate and outage \\
\hline
\cite{liu2022joint, ren-2022RIS} & RIS-assisted SWIPT NOMA & RF PS & Data rate, fairness & RIS+PS improves edge user throughput \\
\hline
\cite{zhang-2022Hybrid} & Hybrid TS/PS with RIS-NOMA & Hybrid RF TS/PS & Outage, diversity gain & Transmit antenna selection + RIS + SWIPT improves diversity \\
\hline
\cite{bajpai-2025Wi, shaikh2024energy, hu2021physical} & UAV-aided SWIPT NOMA & RF PS, WPT & BER, outage, rate & UAV relays improve U2U and cellular links \\
\hline
\cite{mittal2023energy} & UAV-based EH for acoustic NOMA & RF PS & BER, sum-rate & EH UAV boosts underwater relay performance \\
\hline
\cite{dey2024partitioning} & RIS-partitioned EH C-NOMA & WPT, mode switching & EE, throughput & RIS-assisted near-user relaying improves EE \\
\hline
\cite{kumar-2023} & Deep learning-based EH spectrum sharing & RF PS & Outage, throughput, EE & DNN improves EH and UAV spectrum use \\
\hline
\cite{jalali-2024J} & Terahertz SWIPT NOMA for UAVs & RF PS & EE, rate & Power + trajectory tuning boosts EE \\
\hline
\cite{nguyen-2023Se} & UAV-WPT cooperative MEC with NOMA & RF TS & Secrecy rate & PSO optimizes WPT and UAV placement \\
\hline
\cite{shen-2020} & Wireless powered sensor network with NOMA & RF TS & Outage, pairing & UAV-based NOMA extends sensor lifetime \\
\hline
\cite{vo-2020, vo-2021-En} & EH-enabled UAV jammer and relay for IoT & RF TS, dedicated source & Information leakage probability, OP & UAV jamming + EH boosts IoT security \\
\hline
\cite{do-2021} & UAV-NOMA with AF/DF and EH & RF TS & Outage, ergodic capacity & AF/DF UAV relays show EH performance gains \\
\hline
\cite{bhowmick-2021} & Throughput maximization in CR-UAV NOMA & RF EH & Throughput, EE & UAV trajectory + sensing optimized for EH \\
\hline
\end{tabular}
\label{tab:eh_cnoma_lit}
\end{table*}

\subsection{Cognitive Radio and C-NOMA Networks Integration}
Integrating C-NOMA with CR networks is a promising strategy to address the dual challenges of spectrum scarcity and connectivity demands. Many studies have explored C-NOMA-enabled CR networks under various system models, channel conditions, and performance tradeoffs. Early works such as \cite{2019-outage-im, 2020-noma-do, 2018-outage-arzykulov} analyzed outage performance in underlay CR-NOMA systems with imperfect SIC and CSI, establishing analytical baselines and power allocation strategies. Building on this, \cite{2020-underlay-nguyen, m-Chitra-2023} introduced coordinated direct and relay transmission and channel-dependent interference threshold constraints to further enhance reliability and secondary performance. To support energy-constrained deployments, \cite{2021-performance-shukla, bhowmick-2021, 2020-fullduplex-hakimi, 2019-outage-yu} proposed EH-assisted CR-NOMA frameworks. These studies considered RF and hybrid EH along with mobility-aware relaying (e.g., UAV-based), and revealed system throughput and EE gains under practical constraints. Related work in \cite{2024-incremental-p} further introduced incremental relaying with partial relay selection to reduce OP and improve throughput.

Security and reliability tradeoffs were examined in \cite{2018-security-li, 2022-physical-li, 2024-physical-li, garcia-2022-EN}, where physical layer security and secrecy EE were enhanced via cooperative jamming, FD relaying, and AI-driven optimization techniques. \cite{2018-security-li, 2024-physical-li} introduced cooperative jamming and secrecy OP analysis to quantify improvements against eavesdropping. Performance under hardware impairments, loop self-interference, and interference constraints was explored in \cite{arzykulov-2021, 2021-performance-singh, 2019-fullduplex-wang, 2020-hardware-arzykulov, 2018-cooperative-liau}, highlighting the practicality of FD CR-NOMA systems even in non-ideal conditions. Resource allocation, clustering, and DNN-based prediction strategies were employed to optimize throughput and system reliability. \cite{singh-Kumar-2023} further introduced a deep learning model to estimate ergodic sum rate under non-ideal imperfections. \cite{2019-noma-liu} applied cluster-based cognitive IIoT models to optimize sensing and transmission efficiency using NOMA. Overlay and underlay CR-NOMA were further studied in \cite{2021-performance-singh, 2022-cooperative-wang, 2022-performance-agrawal, 2021-performance-yang}, addressing tradeoffs between OP, SE, and system latency under constraints like interference temperature and imperfect SIC. Notably, \cite{2022-cooperative-wang} introduced a hybrid C-NOMA scheme in MEC-assisted CR networks with a two-stage optimization of energy and task allocation, while \cite{2021-performance-singh} proposed an AF cooperative overlay framework accounting for CSI and SIC imperfections. \cite{2019-fundamental-kumar} addressed the limits of spectrum sharing under peak interference constraints in NOMA-based cooperative relaying systems.

Recent studies also integrated RISs and reconfigurable NOMA architectures to enhance cooperative CR networks. For instance, \cite{2024-joint-khan, tin-2024Perf} investigated joint optimization of RIS phase shifts and NOMA power coefficients to boost SE and user fairness in RIS-enabled CR networks. Additionally, \cite{2021-hierarchical-zeng} proposed a hierarchical symbiotic C-NOMA strategy in multicarrier CR networks to support greater connectivity and system throughput. In satellite-integrated CR networks, the work in \cite{chen-2024, zhang-2019, 2021-exploiting-singh, 2020-performance-singh} examined overlay cognitive satellite-terrestrial networks using cooperative C-NOMA and adaptive mode selection, where cooperative relaying improves fairness, SE, and outage performance. UAV-assisted cooperative and cognitive NOMA designs were also explored. The authors in \cite{2021-uav-arzykulov, bhowmick-2021} proposed UAV-based clustering, resource allocation, and flight-aware spectrum-sharing strategies, highlighting the significance of trajectory and sensing optimization in improving throughput and EE. Spectrum leasing and cooperative NOMA with advanced access policies were investigated in \cite{2023-cooperative-hasan}, where a two-phase uplink/downlink scheme with SIC-based decoding improves rate and outage performance compared to traditional OMA approaches. Additionally, advanced techniques like Vickrey auctions \cite{zhang-2020}, deep learning \cite{singh-Kumar-2023}, and symbiotic strategies \cite{2021-hierarchical-zeng} were applied for spectrum access and relay selection. These approaches have further strengthened the flexibility and intelligence of CR-NOMA networks. To summarize, this body of work summarized in Table \ref{tab:crn_cnoma_lit} demonstrates that C-NOMA can significantly enhance CR networks in terms of SE, energy optimization, security, and reliability, especially when augmented with next-generation technologies such as RIS, UAVs, and AI.

\begin{table*}[ht]
\centering
\caption{Summary of Related Work on C-NOMA in Cognitive Radio Networks}
\scalebox{0.95}{
\begin{tabular}{|l|p{5.0cm}|p{1.7cm}|p{2.5cm}|p{5.5cm}|}
\hline
\textbf{Ref.} & \textbf{Scenario / Focus} & \textbf{Access Mode} & \textbf{Metric(s)} & \textbf{Key Insight} \\
\hline
\cite{2019-outage-im,2020-noma-do,2018-outage-arzykulov} & Underlay CR-NOMA with imperfect SIC and CSI & Underlay & OP& Baseline analysis with closed-form expressions and SIC impact \\ 
\hline
\cite{2020-underlay-nguyen, m-Chitra-2023} & CR-NOMA with coordinated/threshold-based power control & Underlay & OP& Power adaptation and coordination improve reliability \\ 
\hline
\cite{2021-performance-shukla,2024-incremental-p} & Overlay CR-NOMA with incremental relaying and EH & Overlay & Outage, EE & AF/DF and partial relay selection strategies reduce OP and enhance EE \\ 
\hline
\cite{bhowmick-2021,2020-fullduplex-hakimi} & UAV/FD CR-NOMA with EH optimization & Overlay & Throughput, rate region & Trajectory and EH model optimize UAV-based CR performance \\ 
\hline
\cite{2019-outage-yu} & SWIPT-enabled cooperative CR-NOMA & Underlay & OP& SWIPT maintains diversity with low OP degradation \\ 
\hline
\cite{2018-security-li,2022-physical-li,2024-physical-li} & Secure CR-NOMA with jamming and SU selection & Overlay & Secrecy OP, Intercept probability & Cooperative jamming and user selection enhance secrecy \\ 
\hline
\cite{garcia-2022-EN} & AI-driven secure FD CR-NOMA & Underlay & Secrecy EE & Ensemble learning + quantum PSO optimize secrecy EE \\ 
\hline
\cite{arzykulov-2021,2020-hardware-arzykulov} & CR-NOMA with hardware impairments & Underlay & OP, SE & Non-ideal hardware and CSI effects analyzed \\ 
\hline
\cite{2021-exploiting-singh,singh-Kumar-2023} & FD overlay CR-NOMA with DNN support & Overlay & Outage, ergodic sum rate & Deep learning predicts ergodic sum rate under impairments \\ 
\hline
\cite{2019-noma-liu} & Cluster-based CR-NOMA for IIoT & Overlay & Throughput & Clustering and resource optimization improve IIoT access \\ 
\hline
\cite{2021-performance-singh} & AF overlay CR-NOMA & Overlay & Outage, diversity & CSI/SIC-aware AF relaying performance \\ 
\hline
\cite{2022-cooperative-wang,2019-fundamental-kumar} & Hybrid CR-NOMA with MEC and SS limits & Hybrid & Rate, OP & Two-stage optimization under peak interference \\ 
\hline
\cite{2022-performance-agrawal} & FD device-to-device-assisted CR-NOMA & Underlay & Throughput & Interference temperature limit-aware allocation for FD relaying \\ 
\hline
\cite{2021-performance-yang} & Dynamic power allocation in CR-NOMA & Underlay & Outage, sum-rate & Adaptive transmission enhances reliability \\ 
\hline
\cite{2024-joint-khan, tin-2024Perf} & RIS-enhanced CR-NOMA systems & Overlay & Sum-rate, EE & Phase/power co-optimization improves fairness \\ 
\hline
\cite{2021-hierarchical-zeng} & Symbiotic CR-NOMA in multicarrier networks & Overlay & Throughput & Hierarchical strategy supports user scaling \\ 
\hline
\cite{zhang-2020} & Cognitive hybrid satellite-terrestrial overlay networks & Overlay & Capacity & Distributed auction improves relay selection \\ 
\hline
\cite{chen-2024,zhang-2019,2021-exploiting-singh,2020-performance-singh} & CR-NOMA in hybrid satellite-terrestrial networks & Overlay/Hybrid & OP, fairness, diversity & Satellite-terrestrial CR networks benefit from cooperative C-NOMA \\ 
\hline
\cite{2023-cooperative-hasan} & Spectrum leasing with cooperative NOMA & Underlay & Rate, OP & Uplink/downlink scheme improves decoding and reuse \\ 
\hline
\end{tabular}
}
\label{tab:crn_cnoma_lit}
\end{table*}

\subsection{Reconfigurable Intelligent Surfaces and C-NOMA Integration}
The integration of RISs into C-NOMA systems has been widely investigated using various operational modes and deployment strategies. In early works such as \cite{wang-2022-RIS, samy-2024Cooperative}, passive fixed RISs were compared with conventional relays, showing competitive performance under high SNR. Other studies such as \cite{gu2024imperfect, zuo-2021Re, gu-2023} explored performance under imperfect SIC, multiple RIS setups, and device-to-device cooperation. Optimization frameworks for passive RIS-based C-NOMA were proposed in \cite{muhammad-2023, liu-2023Optimization, ma-2023Power}, addressing age of information, inter-RIS reflection, and power minimization. Hybrid SWIPT protocols with passive RIS beamforming and antenna selection were analyzed in \cite{zhang-2022Hybrid}.

Active RISs have been increasingly studied to overcome the double fading effect faced by passive RISs and to improve security and signal quality in C-NOMA networks. In \cite{kang-2023Active}, cooperative jamming was integrated with active RIS to enhance physical-layer secrecy. A secure transmission strategy with artificial noise and joint beamforming was proposed in \cite{wang2024active}, where successive convex approximation (SCA) and semi-definite relaxation (SDR) were used to optimize the active RIS and BS configuration. In \cite{wang2024transmission}, a double cooperative active RIS system was studied to mitigate path loss over long distances using majorization-minimization (MM) and fractional programming (FP), achieving significant rate improvement over double passive RIS setups. Active RIS-aided SWIPT was explored in \cite{ren2023transmission}, where the system maximized the weighted sum-rate of users while meeting EH and power constraints. A joint optimization approach with alternating optimization (AO) showed clear advantages over passive schemes.

Secure transmission with a shared power supply between the transmitter and RIS was studied in \cite{dong2022active}, highlighting the benefits of active RISs under realistic power constraints. In uplink scenarios, \cite{peng2024active} proposed an active RIS-aided massive MIMO system using low-resolution analog-to-digital converters (ADCs) and maximum ratio combining (MRC). A genetic algorithm (GA) was used to optimize the RIS phase shifts, with results confirming notable performance improvements over passive RISs. Additionally, the comparative analysis in \cite{nguyen2024performance} offered theoretical insights into OP and ergodic capacity for both active and passive RISs in MISO systems over Nakagami-$m$ fading with imperfect CSI. It was found that active RISs dominate at low and moderate transmit power, though large passive RIS arrays can also yield strong diversity and coding gains.

The development of STAR-RISs introduced bidirectional propagation control. Studies such as \cite{toregozhin2023performance, wang2025sum, singh2024performance, yue-2022Simul} focused on spectral and EE, outage performance, and SIC robustness under STAR-RIS in C-NOMA. The joint use of SWIPT and caching with STAR-RIS was examined in \cite{vu-2023STAR} to address energy-rate tradeoffs.

Most RIS implementations were initially assumed to be fixed on infrastructure. However, UAV-mounted RIS has gained significant attention due to its flexibility and enhanced coverage. In \cite{tyrovolas2022energy}, UAV-mounted RIS was used for energy-efficient IoT data collection, while \cite{tyrovolas2024energy} compared UAV-mounted RIS with UAV-mounted full-duplex relays under trajectory and energy constraints. Secure NOMA with UAV-mounted STAR-RIS was addressed in \cite{guo2023secure}, which incorporated DRL for trajectory control and developed a 3D mobility-aware MIMO channel model to analyze the impact of UAV position variation, while a deployment-element assignment strategy using integer linear programming was developed in \cite{zhao2022joint}. UAV-mounted RIS was also applied in MEC scenarios in \cite{zhai2022energy}, where trajectory, beamforming, and resource allocation were jointly optimized for EE. Orientation uncertainties of UAV-mounted RIS in flight were evaluated in \cite{muller2024assessing} using an extended Kalman filter.

In addition to core throughput and energy targets, physical-layer security and V2X contexts have also motivated RIS-C-NOMA integration. For instance, \cite{ghadi-2024Physical} proposed dual-RIS systems for secure V2V communication under Fisher–Snedecor fading, while \cite{samir-2024Outage} examined ambient backscatter C-NOMA with RIS in V2X setups under imperfect SIC. On-Off-controlled RIS-NOMA networks were addressed in \cite{pei-2023Sec}, focusing on secrecy outage improvements under diverse channel and decoding conditions.

Collectively, these works demonstrate the breadth of RIS-C-NOMA research across passive, active, hybrid, and STAR configurations, with both fixed and UAV-mounted deployments. Table~\ref{tab:ris_cnoma_lit} provides a structured summary of the literature based on these design dimensions.

\begin{table*}[ht]
\centering
\caption{Summary of Related Work on RIS and C-NOMA Integration}
\scalebox{0.85}{
\begin{tabular}{|l|p{4.5cm}|p{2.1cm}|p{2.1cm}|p{3.6cm}|p{5.5cm}|}
\hline
\textbf{Ref.} & \textbf{Scenario/Focus} & \textbf{Deployment Strategy} & \textbf{RIS Mode} & \textbf{Metric(s)} & \textbf{Key Insight} \\
\hline
\cite{wang2024active} & Active RIS with artificial noise for secure downlink transmission & Fixed & Active & Secrecy rate & Joint beamforming and power allocation enhance secrecy via AN and SCA/SDR \\
\hline
\cite{wang2024transmission} & Double cooperative active RIS-aided downlink transmission & Fixed & Active & Achievable rate & Overcomes passive RIS limits; MM/FP-based AO boosts rate under path loss \\
\hline
\cite{nguyen2024performance} & Performance analysis of active vs. passive RIS in MISO under Nakagami-$m$ fading & Fixed & Active & OP, ergodic capacity & Active RIS excels in low/moderate power; better diversity under imperfect CSI \\
\hline
\cite{ren2023transmission} & Active RIS-aided SWIPT with weighted sum-rate maximization & Fixed & Active & Weighted sum-rate, EH & Active RIS mitigates fading; AO-based method ensures EH/QoS \\
\hline
\cite{dong2022active} & Active RIS-aided secure MISO transmission with shared power source & Fixed & Active & Secrecy rate & Shared power design boosts secrecy even with low RIS power \\
\hline
\cite{peng2024active} & Active RIS-aided massive MIMO uplink with low-resolution ADCs & Fixed & Active & Achievable rate & GA-based phase tuning and low-res ADCs show passive RIS gains\\
\hline
\cite{wang-2022-RIS} & RIS vs. relay-assisted C-NOMA & Fixed & Passive & SE & RIS outperforms relay at high SNR; relays better with fewer elements \\
\hline
\cite{samy-2024Cooperative} & RIS-assisted DF relaying in C-NOMA & Fixed & Passive & Coverage, reliability & Outage performance improved using RIS \\
\hline
\cite{gu2024imperfect} & RIS vs. relay under imperfect SIC & Fixed & Passive & Reliability & Relay better unless RIS has many elements \\
\hline
\cite{toregozhin2023performance} & STAR-RIS in C-NOMA & Fixed & STAR-RIS & Spectral and EE & STAR-RIS outperforms AF relay \\
\hline
\cite{wang2025sum} & Sum-rate optimization for STAR-RIS C-NOMA & Fixed & STAR-RIS & Sum-rate & Low-complexity algorithms enhance sum-rate \\
\hline
\cite{singh2024performance} & STAR-RIS vs. AF relay in V2V & Fixed & STAR-RIS & Achievable rate & STAR-RIS performs better at high SNR \\
\hline
\cite{vu-2023STAR} & STAR-RIS with SWIPT and caching & Fixed & STAR-RIS & Energy-rate tradeoff & SWIPT and caching enhance performance \\
\hline
\cite{zuo-2021Re} & Multi-RIS + DF relaying in C-NOMA & Fixed & Passive & Outage, coverage & Improved performance over single-RIS \\
\hline
\cite{gu-2023} & RIS + device-to-device relaying with TDMA & Fixed & Passive & Outage, cooperation & Enhanced cooperation through TDMA reflection \\
\hline
\cite{tyrovolas2024energy} & UAV-mounted RIS vs. UAV-mounted FDR with trajectory optimization & UAV-mounted & Passive & Minimum rate, user fairness, energy & Trajectory optimization boosts fairness and energy use \\
\hline
\cite{muhammad-2023} & RIS-assisted C-NOMA uplink & Fixed & Passive & AoI & Difference-of-convex programming and SCA for optimization \\
\hline
\cite{liu-2023Optimization} & Multi-RIS in multi-cell C-NOMA & Fixed & Passive & Power efficiency & Inter-RIS reflection optimized \\
\hline
\cite{guo2023secure} & Secure uplink NOMA with UAV-mounted STAR-RIS & UAV-mounted & STAR-RIS & Secrecy EE (SEE), data rate & UAV-mounted STAR-RIS improves SEE; DDQN-based trajectory design achieves near-optimal secure performance \\
\hline
\cite{hassan2024optimizing} & Dynamic RIS in mMIMO C-NOMA & UAV-mounted & Dynamic & Scalability, latency & RIS mobility and traffic-aware optimization \\
\hline
\cite{ma-2023Power} & Double cooperative RIS for uplink & Fixed & Passive & Power efficiency & Block coordinate descent algorithm used \\
\hline
\cite{zhang-2022Hybrid} & SWIPT hybrid TS/PS under RIS-NOMA & Fixed & Hybrid & EE & Diversity gain via transmit antenna selection and RIS \\
\hline
\cite{tyrovolas2022energy} & UAV-mounted RIS for IoT data collection & UAV-mounted & Passive & Coverage probability, energy model, throughput & Practical UAV-RIS design improves coverage and EE \\
\hline
\cite{ghadi-2024Physical} & Dual-RIS in V2V NOMA & Fixed & Passive & Physical layer security & Secure transmission using Fisher–Snedecor model \\
\hline
\cite{samir-2024Outage} & Ambient backscatter communication RIS-NOMA in V2X & Fixed & Passive & OP & Imperfect SIC analyzed \\
\hline
\cite{pei-2023Sec} & RIS-NOMA with On-Off control & Fixed & Passive & Secrecy outage & Improved physical layer security, with RIS \\
\hline
\cite{yue-2022Simul} & STAR-RIS-assisted NOMA & Fixed & STAR-RIS & Outage, ergodic rate & Analyzed performance under SIC models \\
\hline
\cite{kang-2023Active} & Active RIS in C-NOMA & Fixed & Active & Security, outage & Active RIS improves performance with cooperative jamming \\
\hline
\cite{guo2023secure} & UAV-mounted RIS-assisted MIMO with 3D mobility & UAV-mounted & Passive & Achievable rate & STAR-RIS and DDQN improve secure performance \\
\hline
\cite{zhao2022joint} & Joint deployment and assignment of multiple UAV-mounted RISs & UAV-mounted & Passive & Number of UAVs, user SNR & ILP-based optimization reduces UAV count while meeting SNR demands across all users \\
\hline
\cite{zhai2022energy} & UAV-mounted RIS-assisted MEC & UAV-mounted & Passive & EE & Joint trajectory, passive beamforming, and MEC resource optimization enhance system EE using SCA and Dinkelbach methods \\
\hline
\cite{muller2024assessing} & EKF-based orientation tracking for UAV-mounted RIS & UAV-mounted & Passive & Orientation accuracy, channel reliability & EKF enables RIS orientation estimation in flight; analysis shows how misalignment degrades RIS link performance \\
\hline
\end{tabular}
}
\label{tab:ris_cnoma_lit}
\end{table*}

\subsection{Space-Air-Ground Integrated Networks and C-NOMA Integration}
C-NOMA has been integrated into space- and air-aided networks to address the demands of 6G-driven SAGIN architectures. In aerial platforms, studies such as \cite{hu-2024,cheepurupalli-2025} incorporated STAR-RIS and MEC into UAV-assisted C-NOMA for improved computation capacity and EE. Security-aware UAV relaying has also been investigated in \cite{nguyen-2023Se,mirbolouk2020relay}, where DF cooperation, coordinated multi-point communication strategies, and jamming mitigation schemes improved secrecy and throughput in dynamic topologies. Hybrid satellite-terrestrial integrated networks (STINs) benefit from relay-aided and spectrum-sharing C-NOMA, as explored in \cite{chen-2024,zhang-2019,singh2020underlay,2020-on-nguyen}. These works addressed SIC-aware cognitive access under realistic fading, while multi-layer and regenerative relay configurations in \cite{ge-2021,wang2023performance,li-2020} enhanced coverage and flexibility for space-ground user distribution. Beamforming and multibeam coordination challenges in satellite NOMA were tackled in \cite{beigi2018capacity,beigi2018interference,yeom2023mmse,li2021cooperative,toka2022outage}, where MIMO and asynchronous reception schemes offered enhanced spatial separation and capacity despite CSI imperfections and beam overlaps.

To meet stringent QoS and delay requirements in LEO-based STINs, learning- and game-theoretic optimizations were introduced in \cite{gao-2021,li-2024,shen-2024,2022-non-gao,2021-max-gao}, optimizing subcarrier pairing, age of information (AoI), and bitrate adaptation across physical and medium access control (MAC) layers. Security and resilience in satellite IoT (SatIoT) networks are strengthened by cooperative FD-NOMA \cite{2019-secrecy-yin}, anti-jamming grouping \cite{2023-anti-han}, and packet repair via automatic repeat request with terrestrial feedback \cite{2024-integrated-ahmed}, forming a robust protection stack against eavesdropping and denial-of-service attacks. Advanced access methods, such as rate-splitting multiple access (RSMA) and auction-based relay selection were proposed in \cite{2025-realistic-li,2024-joint-han,zhang-2020} to improve user fairness and relay coordination under interference-limited and shared-spectrum settings. For multimedia broadcast applications, chunk-level distortion-aware NOMA \cite{2022-cooperative-zhang} and symbol error rate-driven optimization under hardware impairments \cite{2024-symbol-xu} ensured reliable service delivery over heterogeneous SAGIN nodes. In a nutshell, these works concluded in Table \ref{tab:cnoma_sagin_lit} underscore the versatility of C-NOMA across SAGIN layers and highlight ongoing challenges in scaling SIC reliability, mitigating hardware nonidealities, and coordinating access under partial CSI, making it a cornerstone of future cooperative 6G networks.

\begin{table*}[ht]
\centering
\caption{Summary of Related Work on C-NOMA in Space-Air-Ground Integrated Networks (SAGIN)}
\scalebox{0.95}{
\begin{tabular}{|l|p{4.9cm}|p{2.9cm}|p{6.0cm}|}
\hline
\textbf{Ref.} & \textbf{Application Scenario} & \textbf{Metric(s)} & \textbf{Key Insight} \\
\hline
\cite{hu-2024,cheepurupalli-2025} & UAV-assisted C-NOMA with MEC and STAR-RIS & EE, computation rate & Joint UAV-RIS design enhances offloading and energy metrics \\
\hline
\cite{nguyen-2023Se,mirbolouk2020relay} & UAV-assisted relay and security-aware C-NOMA & Secrecy outage, sum rate & Relaying and jamming enhance security and spectrum use \\
\hline
\cite{chen-2024,zhang-2019,singh2020underlay,2020-on-nguyen} & Cognitive HSTN with relay-assisted NOMA & OP, ergodic capacity & Hybrid relaying with SIC boosts SE \\
\hline
\cite{ge-2021,wang2023performance,li-2020} & Layered and coordinated satellite NOMA & BER, OP & Multi-layer relays extend user reliability and coverage \\
\hline
\cite{beigi2018capacity,beigi2018interference,li2021cooperative,toka2022outage,yeom2023mmse} & Multibeam, beam hopping, and MIMO satellite NOMA & Sum rate, BER, outage & Beam coordination and SIC enhance capacity and decoding \\
\hline
\cite{gao-2021,li-2024,shen-2024,2022-non-gao,2021-max-gao} & QoE and latency-aware NOMA in LEO STINs & AoI, QoE, latency, completion time & Optimization frameworks improve delay and user satisfaction \\
\hline
\cite{2019-secrecy-yin,2023-anti-han,2024-integrated-ahmed} & Security and robustness in SatIoT-NOMA & Secrecy rate, packet drop rate, throughput & Cooperative and anti-jamming schemes strengthen resilience \\
\hline
\cite{2025-realistic-li,2024-joint-han,zhang-2020} & Advanced access: RSMA, auctions, spectrum sharing & Max-min rate, capacity & Hybrid access models improve fairness and allocation \\
\hline
\cite{2022-cooperative-zhang,2024-symbol-xu} & Multimedia multicast and symbol-level metrics & Video distortion, symbol error rate & Robust coding and DF relays combat interference and impairments \\
\hline
\end{tabular}
\label{tab:cnoma_sagin_lit}
}
\end{table*}

\subsection{Integrated Sensing and Communication and C-NOMA Integration}
The integration of C-NOMA with ISAC is still in its infancy, but it represents a transformative step toward spectrum- and energy-efficient wireless systems. By enabling simultaneous information transfer and environmental sensing, ISAC can benefit significantly from the SE, user fairness, and diversity gains offered by C-NOMA. One of the earliest studies in this space, \cite{2024-covert-li}, proposed a covert beamforming strategy for a cooperative NOMA-assisted ISAC system, focusing on secure target detection while ensuring information reliability. The work addressed the optimization of sensing beampatterns and power control under covertness constraints. Similarly, \cite{2025-transmit-wang} investigated transmit power optimization for secure C-NOMA-based ISAC systems with SWIPT, leveraging convex approximations to maximize sum-rate and sensing SINR.

The authors in \cite{2023-integrated-amhaz} provided a comparative analysis of conventional NOMA versus C-NOMA in an ISAC setting, showing that cooperation among users improves both sensing accuracy and communication coverage under imperfect SIC conditions. Building on that, \cite{2024-uav-amhaz} integrated UAVs and DDPG algorithms to optimize the trajectory and power control in a C-NOMA-enabled ISAC framework, revealing significant sensing accuracy and SE improvements. In the context of secure multicast and unicast communications, \cite{2024-secure-zhang} investigated beamforming design in NOMA-ISAC networks. This work demonstrated how the multicast and unicast messages can be simultaneously transmitted with optimal resource allocation while satisfying sensing constraints. Moreover, \cite{2024-rendezvous-nasser} provided a comprehensive survey on the convergence of NOMA and ISAC, outlining current trends, technical enablers, and future research directions, such as joint waveform design, user clustering, and secure ISAC. These works establish the potential of C-NOMA-ISAC systems to unlock new opportunities for joint connectivity and situational awareness in 6G and beyond networks summarized in Table \ref{tab:cnoma_isac_lit}. However, further studies are needed to develop unified frameworks that optimize both sensing and communication objectives under diverse user, mobility, and spectrum constraints.

\begin{table*}[ht]
\centering
\caption{Summary of Related Work on C-NOMA and ISAC Integration}
\begin{tabular}{|l|p{5.8cm}|p{4.0cm}|p{5.9cm}|}
\hline
\textbf{Ref.} & \textbf{Scenario/Focus} & \textbf{Metrics} & \textbf{Key Insight} \\
\hline
\cite{2024-covert-li} & C-NOMA-assisted ISAC with covert beamforming & Secrecy rate, detection probability & Joint design of secure sensing and communication beamforming \\
\hline
\cite{2025-transmit-wang} & C-NOMA-based ISAC with SWIPT & Sum-rate, sensing SINR & Convex-based optimization of transmit power and sensing quality \\
\hline
\cite{2023-integrated-amhaz} & NOMA vs. C-NOMA in ISAC & Sensing accuracy, coverage & C-NOMA provides higher accuracy under imperfect SIC \\
\hline
\cite{2024-uav-amhaz} & UAV-assisted C-NOMA for ISAC using DDPG & Sensing accuracy, SE & DDPG learns UAV trajectory and power control policy \\
\hline
\cite{2024-secure-zhang} & Secure multicast/unicast beamforming for NOMA-ISAC & Information rate, sensing constraint & Joint multicast-unicast transmission under security and sensing QoS \\
\hline
\cite{2024-rendezvous-nasser} & Survey on ISAC-NOMA integration & N/A & Roadmap for waveform design, user pairing, and secure ISAC integration \\
\hline
\end{tabular}
\label{tab:cnoma_isac_lit}
\end{table*}

\subsection{Optimization in C-NOMA Networks}
With the increasing complexity of C-NOMA systems, particularly when integrated with emerging technologies such as cognitive radio, RIS, EH, ISAC, and satellite-terrestrial networking, optimization has become central to performance enhancement. The literature reveals a variety of technique, ranging from model-based formulations to heuristic and AI-driven approaches, applied to solve challenges such as power allocation, user pairing, relay selection, trajectory design, beamforming, task offloading, and resource scheduling, subject to practical constraints like transmit power budgets, EE, QoS, secrecy outage, interference tolerance, and fairness. We classify the most recent state-of-the-art optimization works into model-based, heuristic and metaheuristic, and AI-based as follows.

\textbf{Model-based optimization} remains the dominant approach in the literature, especially for convex and tractable system formulations. Works like \cite{2025-transmit-wang} applied convex approximation methods for transmit power optimization in secure ISAC-C-NOMA with SWIPT, subject to secrecy rate and total power constraints. Similarly, \cite{2020-noma-do, 2019-fundamental-kumar, 2021-performance-yang, li-2020Adaptive, mittal2023performance, karthik2021cooperative, thakur2021novel, sarkar2022outage, neole2024dynamic, ghosh2022energy, bhowmick-2021} studied power allocation and outage minimization in EH-based and cooperative relay-aided C-NOMA networks. \cite{2022-cooperative-wang} formulated a two-stage energy consumption minimization problem for MEC-aided cognitive NOMA networks by jointly allocating power, time slots, and computation tasks. \cite{2021-overlay-singh} investigated power allocation for EH and relay operation using TS/PS protocols. Difference-of-convex programming and SCA were employed in \cite{muhammad-2023, zhai-2024Resource, yang-2023Delay, ren-2022RIS, fu-2024Sum, elhattab-2021, chen-2024Coop, khisa-2023, qin-2025J, zuo-2021Re, hu-2024, liu2022joint, gao-2021, 2021-max-gao, 2024-joint-khan, 2021-uav-arzykulov, 2020-fullduplex-hakimi} to jointly optimize transmit powers, RIS beamforming vectors, PS/TS ratios, central processing unit (CPU) frequencies, and NOMA power coefficients under various QoS and system design constraints. Works like \cite{wang2025sum, ma-2023Power, jalali-2024J, li2021cooperative, toka2022outage, 2025-realistic-li, 2022-cooperative-zhang, 2019-noma-liu, 2021-performance-shukla, 2021-performance-yang} targeted sum-rate, EE, throughput, or fairness, employing fractional programming, mixed-integer programming, or alternating optimization techniques. The authors in \cite{2021-hierarchical-zeng, 2023-cooperative-hasan, 2023-cooperative-hasan, 2022-performance-agrawal} included strategies for joint resource optimization in multi-carrier and spectrum-leasing C-NOMA systems. The paper in \cite{arzykulov-2021} proposed a linear bottleneck assignment-based algorithm for UAV-based resource and channel allocation under CSI uncertainty. 

\textbf{Heuristic and metaheuristic optimization} have also gained traction, particularly for highly non-convex and large-scale optimization problems. \cite{garcia-2022-EN} proposed a bi-level optimization framework combining ensemble learning for relay selection and quantum-behaved particle swarm optimization (QPSO) for power allocation to maximize secrecy EE. \cite{2024-incremental-p} applied a partial relay selection scheme under incremental relaying to optimize outage and throughput tradeoffs. \cite{zhang-2020} utilized a Vickrey auction-based scheme to select the optimal cooperative relay in hybrid satellite-terrestrial overlay networks. \cite{garcia-2022Low} employed PSO to minimize transmission power and maximize EE in SWIPT-enabled NOMA systems under nonlinear EH models. \cite{nguyen-2023Se} leveraged PSO for UAV positioning and secrecy enhancement under multi-objective constraints. \cite{mirbolouk2020relay} optimized relay selection under SINR and sum-rate constraints in satellite coordinated multi-point NOMA. \cite{zhao-2023} solved a fairness-aware rate optimization problem under time-asynchronous transmission using Gale-Shapley matching and a Dinkelbach-like power allocation algorithm. \cite{2023-performance-chai} optimized power allocation for mean EE in hybrid satellite-terrestrial NOMA with co-channel interference, while \cite{2021-noma-shuai} used partial relay selection to balance outage and complexity under imperfect SIC.

\textbf{AI-based optimization} has emerged as a promising solution for handling high-dimensional, imperfect, and dynamic environments \cite{chandra2023multi}. \cite{singh-Kumar-2023} developed a DNN framework for ergodic secrecy prediction in FD CR-NOMA systems under hardware impairments and SIC imperfections. \cite{2024-uav-amhaz, li-2024} used DDPG to jointly optimize trajectory and power for UAV-assisted ISAC and satellite multicast systems with constraints on energy use, QoE, and fairness. \cite{garcia-2022-EN} further applied ensemble learning to reduce the search space in relay assignment, enabling faster convergence. Simulation-based policy tuning and learning frameworks such as DRL, DQN, and online adaptation strategies were employed in \cite{hassan2024optimizing, gu2024imperfect, kang-2023Active, qin-2024Deep, cheepurupalli-2025, kumar-2023, tran-2022, shen-2024} to jointly manage transmit power, RIS elements, secrecy constraints, and resource allocation strategies under realistic hardware impairments and channel uncertainty.

In summary, the diverse use of model-based, heuristic, and AI-based techniques in the literature reflects the growing need for scalable and adaptive solutions in C-NOMA networks. Future systems will likely benefit from hybrid frameworks that combine theoretical tractability with learning-based adaptability to achieve real-time and robust performance under practical constraints.

\begin{table*}[ht]
\centering
\caption{Summary of Optimization Techniques in C-NOMA Literature}
\scalebox{0.99}{
\begin{tabular}{|p{4.7cm}|p{5.5cm}|p{7cm}|}
\hline
\textbf{Ref.} & \textbf{Category / AI Technique} & \textbf{Optimization Objective(s)} \\
\hline
\cite{2025-transmit-wang, 2020-noma-do, 2019-fundamental-kumar} & Model-Based & Transmit power minimization, secrecy rate maximization \\
\hline
\cite{2022-cooperative-wang, 2021-overlay-singh} & Model-Based & Energy consumption minimization, relay resource allocation \\
\hline
\cite{muhammad-2023, yang-2023Delay, ren-2022RIS, fu-2024Sum} & Model-Based & RIS phase optimization, SWIPT ratio tuning \\
\hline
\cite{khisa-2023, qin-2025J, zuo-2021Re, hu-2024} & Model-Based & Joint CPU and PS/TS tuning, QoS-aware power control \\
\hline
\cite{zhai-2024Resource, elhattab-2021} & Model-Based & Beamforming, relay power allocation \\
\hline
\cite{li2021cooperative, toka2022outage, jalali-2024J} & Model-Based & Sum-rate maximization, EE optimization \\
\hline
\cite{arzykulov-2021, 2024-joint-khan, 2021-performance-yang} & Model-Based & Channel/resource allocation, user clustering \\
\hline
\cite{garcia-2022-EN} & Metaheuristic + machine learning (QPSO + ensemble learning)& Relay selection, power allocation \\
\hline
\cite{2024-incremental-p} & Heuristic & Incremental relaying, outage throughput balance \\
\hline
\cite{zhang-2020} & Heuristic & Relay auction-based selection \\
\hline
\cite{garcia-2022Low} & Heuristic (PSO) & Transmission power and EH efficiency \\
\hline
\cite{nguyen-2023Se} & Heuristic (PSO) & UAV trajectory and secrecy rate \\
\hline
\cite{mirbolouk2020relay} & Heuristic & Relay selection in satellite coordinated multi-point \\
\hline
\cite{zhao-2023} & Heuristic (Matching + Dinkelbach) & Rate fairness with time-asyncronous  design \\
\hline
\cite{2023-performance-chai, 2021-noma-shuai} & Heuristic & Power control under imperfect SIC \\
\hline
\cite{singh-Kumar-2023} & AI-based (DNN) & Ergodic secrecy rate prediction under impairments \\
\hline
\cite{2024-uav-amhaz, li-2024} & AI-based (DDPG) & Trajectory and power in UAV-assisted networks \\
\hline
\cite{garcia-2022-EN} & AI-based (Ensemble learning) & Relay selection via learning \\
\hline
\cite{hassan2024optimizing, gu2024imperfect, kang-2023Active, qin-2024Deep} & AI-based (DRL algorithms such as DQN) & Joint control under hardware uncertainty \\
\hline
\end{tabular}
}
\label{tab:optimization_cnoma}
\end{table*}

\section{Challenges and Open Problems} \label{Sec_chall}
\subsection{C-NOMA vs Energy Harvesting}
EH in C-NOMA systems, particularly for UAVs and edge devices with limited resources, presents significant challenges. Although C-NOMA can boost EE by increasing harvested energy, it still relies on antenna design and the amount of harvested energy. Therefore, a more sophisticated antenna design should balance EH efficiency and communication performance. Besides, combining RF and renewable energy sources can be a feasible solution to satisfy the QoS requirements.  Additionally, lightweight AI models are crucial for managing energy in low-power devices. Future efforts should prioritize advancements in WPT, ambient EH, and backscattering technologies. Distributed AI can optimize energy collection from inconsistent sources, while large-scale AI-driven studies are vital for jointly optimizing EH and resource allocation, ensuring EE in ultra-dense networks.

\subsection{C-NOMA vs Reconfigurable Intelligent Surfaces}
RISs, as a low-cost array of passive reflecting elements, can significantly boost C-NOMA performance by enabling large-scale cooperative MIMO-NOMA networks. However, several challenges remain. Designing optimal reflecting coefficients based on user locations and channel conditions is complex, especially in the absence of active RF chains. This brings near-field communication into play, where RISs have to estimate both CSI and visibility regions to effectively direct signals. Environmental factors like dust and humidity can degrade performance, while signal redirection may increase interference for others. AI-driven optimization is essential to dynamically adjust RIS elements, ensuring optimal phase, magnitude settings, and efficient power allocation while addressing NOMA pair formation and near-field constraints.

\subsection{C-NOMA vs ISAC-assisted Semantic Communication}
Integrating C-NOMA with ISAC for semantic communication introduces several challenges. While DNNs are effective for feature extraction, they often lack stability in dynamic environments, impacting real-time applications like immersive XR and digital twins. Key open problems include developing lightweight, optimizing DNN architectures for real-time adaptability, and minimizing latency while ensuring high data fidelity. 
\subsection{C-NOMA vs Space-Air-Ground Integrated Networks}
Integrating C-NOMA into SAGINs faces challenges like diverse channel conditions and interference due to the network's varying layers including satellites, UAVs, and ground. While C-NOMA improves SE, it requires adaptive beamforming and power control to handle dynamic topologies. AI-driven optimization can enhance resource allocation and interference management in real-time. Future work should focus on hybrid approaches combining C-NOMA with cooperative relaying and satellite backhaul to boost connectivity and efficiency in SAGINs.

\subsection{C-NOMA vs Digital Twin and Haptic Communications}
Haptic communications and digital twins are essential for applications requiring ultra-low latency and high reliability such as telemedicine. Challenges include maintaining real-time responsiveness through network coordination, resource optimization, and AI-enhanced signal alignment. A key issue with C-NOMA is that the SIC process can impact real-time operations, necessitating more intelligent and adaptive approaches such as native AI to handle dynamic network conditions. Additionally, achieving ultra-low latency in semantic data processing and mitigating signal distortion in high-mobility environments remain open problems. Future research should prioritize adaptive and intelligent solutions for beamforming and Doppler shift mitigation. These solutions should optimize C-NOMA frameworks to enhance the quality and efficiency of instantaneous communications.

\subsection{C-NOMA vs Quantum Communication}
The integration of C-NOMA with quantum communication, particularly quantum key distribution, presents unique challenges due to the sensitivity of quantum states to interference and power variations inherent in NOMA. SIC operations and user multiplexing in C-NOMA can disrupt the fragile coherence required in quantum key distribution, making joint operation difficult without cross-layer adaptation. Moreover, combining semantic communication with quantum links introduces issues in aligning compression efficiency with quantum fidelity constraints. Real-time C-NOMA functions like pairing and beamforming are also limited by the resource constraints of quantum hardware. To overcome these challenges, hybrid classical-quantum resource management and interference-aware scheduling schemes are needed. AI-driven approaches, including reinforcement learning, can optimize coexistence. Future work should focus on lightweight quantum-safe protocols, semantic-aware encoding, and the design of robust C-NOMA frameworks tailored for quantum-enhanced networks.
\subsection{C-NOMA vs Security and Privacy}
As C-NOMA integrates with increasingly intelligent and layered architectures, such as RIS-assisted environments, SAGINs, ISAC, and semantic communication, ensuring secure and private transmission becomes a critical challenge. C-NOMA’s reliance on power-domain user multiplexing and SIC can expose weak users to potential eavesdropping, especially when CSI is imperfect or outdated. Moreover, the integration of semantic communication introduces new risks such as semantic-level inference attacks and context manipulation. In RIS-aided systems, malicious actors may attempt to reprogram passive surfaces or exploit channel reflections to disrupt or intercept transmissions. In SAGINs, the open nature of UAV and satellite links further magnifies these vulnerabilities due to their wide-area exposure and variable latency. Protecting semantic and sensed data in ISAC frameworks requires lightweight encryption and trust mechanisms that preserve real-time performance. Future research should explore secure C-NOMA frameworks that combine physical-layer security with cryptographic enhancements and AI-driven anomaly detection. Federated learning and distributed authentication mechanisms may also help protect user privacy and communication integrity across dynamic, multi-tier networks.

\subsection{Standardization Outlook for Integrated C-NOMA Technologies}
Despite its demonstrated advantages in SE and user fairness, C-NOMA has not been formally adopted into 5G NR standards, primarily due to unresolved challenges in SIC reliability, signaling overhead, and complexity in practical deployments. These challenges become even more pronounced when C-NOMA is integrated with emerging technologies such as RIS, UAV-assisted relaying, massive MIMO, EH, and semantic ISAC. For instance, integrating RIS with C-NOMA introduces additional signaling demands to coordinate passive or hybrid reflecting elements with power-domain multiplexing, while UAV-assisted C-NOMA links require dynamic CSI acquisition under mobility and fluctuating link quality.

Standardization efforts must therefore extend beyond basic NOMA signaling to support heterogeneous architectures, including distributed EH constraints, real-time semantic traffic prioritization, and RIS control granularity. Uniform models for joint CSI feedback, RIS phase configuration, and low-latency control signaling are essential to enable consistent performance under varying deployment scenarios. Furthermore, integrating C-NOMA with ISAC or semantic layers requires rethinking traditional physical-MAC splits to accommodate higher-level feature-awareness in scheduling and decoding. As 5G-Advanced evolves toward 6G, these requirements call for cross-layer co-design and harmonized reference models that support intelligent, adaptive, and collaborative multiple access in highly dynamic environments.

Addressing these challenges with innovative AI techniques, dynamic optimization, and efficient resource management will be essential to fully realize the potential of C-NOMA and its integration with next-generation technologies, ultimately enhancing the QoE for diverse applications in future communication systems.

\section{Conclusion} \label{Sec_Conc}
In this work, we presented a comprehensive and unified survey on the integration of C-NOMA with key enabling technologies envisioned for 6G wireless networks. We began by introducing the fundamental concepts of C-NOMA and its associated relaying protocols, establishing the foundation for understanding its practical relevance and performance advantages in next-generation systems. This was followed by an in-depth investigation of how C-NOMA can be functionally and architecturally integrated with emerging technologies. Beyond technical integration, the survey emphasized the orchestration of C-NOMA across future application domains, where low latency, adaptability, and scalability are essential. A structured literature review was also conducted, highlighting system models, relaying strategies, protocol-level interactions, and performance metrics across these domains. Additionally, a dedicated subsection categorized recent optimization approaches into model-based, heuristic, and AI-driven strategies, underscoring the growing importance of intelligent and adaptive resource management in dynamic environments. Finally, the article identified critical research challenges concerning C-NOMA. Although this work offers extensive coverage of C-NOMA integration, further exploration of areas such as task scheduling, distributed learning coordination, and lightweight security protocols would help enrich the C-NOMA research landscape and accelerate its readiness for real-world deployment.

\bibliographystyle{IEEEtran}
\bibliography{Main}


 





\end{document}